\documentclass[twocolumn]{aastex631}

\newcommand{\upperrad}{$5.8^{+4.8}_{-2.1}R_\oplus$}
\newcommand{\lowerrad}{$4.2^{+3.5}_{-1.6}R_\oplus$}

\shorttitle{PINES II}
\shortauthors{Tamburo et al.}
\graphicspath{{./}{figures/}}

\begin{document}

\title{The Perkins INfrared Exosatellite Survey (PINES) II. Transit Candidates and Implications for Planet Occurrence around L and T Dwarfs}

\correspondingauthor{Patrick Tamburo}
\email{tamburop@bu.edu}

\author[0000-0003-2171-5083]{Patrick Tamburo}
\affiliation{Department of Astronomy \& The Institute for Astrophysical Research, Boston University, 725 Commonwealth Ave., Boston, MA 02215, USA}

\author[0000-0002-0638-8822]{Philip S. Muirhead}
\affiliation{Department of Astronomy \& The Institute for Astrophysical Research, Boston University, 725 Commonwealth Ave., Boston, MA 02215, USA}

\author[0000-0003-2015-5029]{Allison M. McCarthy}
\affiliation{Department of Astronomy \& The Institute for Astrophysical Research, Boston University, 725 Commonwealth Ave., Boston, MA 02215, USA}

\author[0000-0002-0330-1648]{Murdock Hart}
\affiliation{Department of Astronomy \& The Institute for Astrophysical Research, Boston University, 725 Commonwealth Ave., Boston, MA 02215, USA}
\affiliation{Perkins Telescope Observatory, 2780 Marshall Lake Rd., Flagstaff, AZ 86005, USA}

\author[0000-0003-0489-1528]{Johanna M. Vos}
\affiliation{American Museum of Natural History, 200 Central Park West, New York, NY 10024, USA}

\author[0000-0002-0802-9145]{Eric Agol}
\affiliation{Department of Astronomy \& Virtual Planetary Laboratory, University of Washington, Box 351580, U.W., Seattle, WA 98195, USA}

\author[0000-0002-9807-5435]{Christopher Theissen}
\altaffiliation{NASA Sagan Fellow}
\affiliation{Center for Astrophysics and Space Studies, University of California, San Diego, La Jolla, CA 92093-0424, USA}

\author[0000-0002-0447-7231]{David Gracia}
\affiliation{Department of Astronomy \& The Institute for Astrophysical Research, Boston University, 725 Commonwealth Ave., Boston, MA 02215, USA}

\author[0000-0001-8170-7072]{Daniella C. Bardalez Gagliuffi}
\affiliation{Department of Physics \& Astronomy, Amherst College, 25 East Drive, Amherst, MA 01003, USA}
\affiliation{American Museum of Natural History, 200 Central Park West, New York, NY 10024, USA}

\author[0000-0001-6251-0573]{Jacqueline Faherty}
\affiliation{American Museum of Natural History, 200 Central Park West, New York, NY 10024, USA}



\begin{abstract}
We describe a new transit detection algorithm designed to detect single transit events in discontinuous Perkins INfrared Exosatellite Survey (PINES) observations of L and T dwarfs. We use this algorithm to search for transits in 131 PINES light curves and identify two transit candidates: 2MASS J18212815+1414010 (2MASS J1821+1414) and 2MASS J08350622+1953050 (2MASS J0835+1953). We disfavor 2MASS J1821+1414 as a genuine transit candidate due to the known variability properties of the source. We cannot rule out the planetary nature of 2MASS J0835+1953's candidate event and perform follow-up observations in an attempt to recover a second transit. A repeat event has yet to be observed, but these observations suggest that target variability is an unlikely cause of the candidate transit. We perform a Markov chain Monte Carlo simulation of the light curve and estimate a planet radius ranging from $4.2^{+3.5}_{-1.6}R_\oplus$ to $5.8^{+4.8}_{-2.1}R_\oplus$, depending on the host's age. Finally, we perform an injection and recovery simulation on our light curve sample. We inject planets into our data using measured M dwarf planet occurrence rates and attempt to recover them using our transit search algorithm. Our detection rates suggest that, assuming M dwarf planet occurrence rates, we should have roughly a 1\% chance of detecting a candidate that could cause the transit depth we observe for 2MASS J0835+1953. If 2MASS J0835+1953 b is confirmed, it would suggest an enhancement in the occurrence of short-period planets around L and T dwarfs in comparison to M dwarfs, which would challenge predictions from planet formation models.
\end{abstract}



\section{Introduction} \label{sec:intro}

The Perkins INfrared Exosatellite Survey (PINES) is an ongoing near-infrared (NIR) search for transiting planets and moons around a sample of 393 spectroscopically confirmed L and T dwarfs \citep{Tamburo2022}. These spectral types (canonically defined in \citet{Kirkpatrick1999a} and \citet{Burgasser2006}, respectively) span the boundary between the hydrogen-fusing regime of stars and the non-H-fusing regime of brown dwarfs and planetary-mass objects. Due to their diminutive sizes ($\sim$1 R$_{Jup}$) and cool temperatures ($\sim700-2200$ K), L and T dwarfs are faint at optical wavelengths, and previous optical exoplanet surveys have been insensitive to the detection of short-period ($P \lesssim 200 $ days) planetary companions \citep{Sagear2020, Sestovic2020}. PINES, which is performed in $J$- and $H$-bands, and the Search for habitable Planets EClipsing ULtra-cOOl Stars \citep[SPECULOOS;][]{Delrez2018, Murray2020, Sebastian2021}, which operates in the red-optical ($0.7-1.1$ $\mu$m), have demonstrated the photometric precision to detect short-period super-Earth planets around these objects.

The question of whether or not such planets should exist is an active area of research. Kepler discoveries reveal an anti-correlation in the occurrence rate of short-period, super-Earth-sized planets ($1-4$ $R_\oplus$) with host mass, with M dwarfs hosting $\sim3$ times as many such planets as F dwarfs \citep{Howard2012, Dressing2015,Mulders2015a,Mulders2015b}. Kepler data also provides tentative evidence for an enhancement of short-period planet occurrence within the M spectral type itself, with a measured increase in the occurrence of $0.5-2.5$ $R_\oplus$ planets over the range of M3 to M5 \citep{HardegreeUllman2019}. The observed anti-correlation of short-period super-Earth occurrence rates with host mass has been linked to the \textit{positive} correlation of substructure occurrence rates in protoplanetary disks with stellar mass: disks around low-mass stars are, on average, less likely to possess sufficient gas mass to form giant planets in the outer disk, and are therefore less likely to develop substructures that inhibit millimeter grain drift to the inner disk, where super-Earths are thought to form \citep{vanderMarel2021}. Together, these results point to an increased efficiency of smaller stars in forming and/or retaining Earth- to super-Earth-sized planets, and these trends may continue into the L and T spectral types.

However, planet formation simulations and brown dwarf disk mass measurements paint a more complicated picture. A simulation from \citet{Payne2007} suggests that Earth-sized planets should not be able to form around brown dwarfs with less than a Jupiter mass in their protoplanetary disks, and brown dwarf disk mass measurements are frequently lower than this value \citep[e.g., ][]{Rilinger2021}. A pebble accretion model from \citet{Mulders2021}, which captures measured trends in the occurrence rates of giant and super-Earth planets, predicts that super-Earths should not be able to form around hosts in the L and T dwarf mass regime.

To summarize, planet occurrence rate trends point to a sizeable population of short-period super-Earths around L and T dwarfs, while simulations and disk mass measurements do not. Differentiating between these predictions requires searching for such planets around the L and T spectral types. In this work, we perform a systematic search for transiting super-Earths around L and T dwarfs using more than two years of PINES observations. We describe the sample of 131 light curves used in our study in Section \ref{sec:observations}. Our observing cadence renders traditional transit detection algorithms ineffective, and we describe a new algorithm that we have developed to detect single, incomplete transit events in potentially variable data in Section \ref{sec:transit_search_and_recovery_algorithm}. In Section \ref{sec:transit_search}, we report on the results of our transit search, which identified two candidate transit events. Finally, in Section \ref{sec:injection_recovery}, we perform an injection and recovery simulation using our light curve sample and transit detection algorithm. We use this simulation to ask whether or not our data is consistent with measured M dwarf short-period planet occurrence rates.

\section{Observations} \label{sec:observations}
We performed our analysis on PINES light curves that were obtained between May 2020 and March 2022. Observations were taken with the Mimir instrument \citep{Clemens2007} on Boston University's 1.8-m Perkins Telescope Observatory (PTO). Light curves were created with the \texttt{PINES Analysis Toolkit} \citep[\texttt{PAT,}][]{PINES_analysis_toolkit}, which was designed to minimize systematic effects that are unique to PINES data, most notably a large number of bad pixels on the detector \citep{Tamburo2022}. To correct changes in the raw flux of our targets due to evolving observing conditions, the pipeline uses an artificial light curve (ALC), which is a unitless weighted sum of normalized reference star fluxes \citep{Murray2020}. Typical fields contain dozens of suitable reference stars for creating the ALC, thanks to the $10'\times10'$ field-of-view of Mimir data in its wide-field imaging mode. 

Multiple light curves are created for each PINES target in \texttt{PAT}, corresponding to different fixed and time-variable aperture sizes. The radii of the time-variable apertures are set by a multiplicative factor times a smoothed trend of the measured seeing full-width half max (FWHM) values in each image, which is computed using a running average with a width of five points. We selected the light curve that minimized the average scatter over the duration of individual blocks of data for each target, where a \textit{block} refers to the set of exposures that are obtained roughly once per hour for each target during PINES observations. A typical block's duration ranges from 8 to 14 minutes, depending on the number of targets being observed on a given night (see Section \ref{sec:transit_search_and_recovery_algorithm}). 

We then trimmed poor weather nights from the observations, which can result in anomalously variable light curves due to low source counts or inaccurate centroiding. We identified poor weather nights using each target's ALC. We discarded any night whose ALC had a standard deviation greater than $20$\%, which removed 53 out of 430 nights of data. 

The 131 light curves that were included in this analysis, consisting of 377 individual nights of data, are summarized in Table \ref{tab:observations}. Light curves range in duration from 1 to 11 nights, with a median duration of 3 nights. Observations were taken in Mauna Kea Observatories (MKO) $J$- and $H$-bands \citep{Simons2002,Tokunaga2002}, which \citet{Tamburo2022} found are relatively insensitive to systematic effects from variable precipitable water vapor (PWV). A total of 96 light curves were taken in $J$-band, while 35 were taken in $H$-band. Calculations from \citet{Tamburo2019} suggested that $H$-band would provide the highest signal-to-noise ratio (SNR) data on our targets, but their study assumed an average seeing FWHM of $1.^{\prime\prime}5$. We have measured an average seeing FWHM of $2.^{\prime\prime}6$ through the first 20 months of the survey \citep{Tamburo2022}, which necessitates larger aperture sizes with higher contributions from sky background. Under these conditions, we find that $J$-band offers the highest SNR data, and almost all observations since May 2020 have been performed in $J$-band as a result.

\begin{deluxetable*}{lllllllrrr}
    \tablenum{1}
    \tablecaption{PINES observations used in this study. The full table is available online.}
    \label{tab:observations}
    \tablewidth{0pt}
    
    \tablehead{
        \colhead{2MASS} & \colhead{R.A.}    & \colhead{Dec.}    & \colhead{2MASS} & \colhead{SpT}  & \colhead{SpT} & \nocolhead{}   & \nocolhead{}           & \colhead{Exposure} & \colhead{$\sigma$} \\
        \colhead{Name}  & \colhead{(J2000)} & \colhead{(J2000)} & \colhead{$J$} & \colhead{Opt.} & \colhead{NIR} & \colhead{Band} & \colhead{$N_{nights}$} & \colhead{Time (s)} & \colhead{(\%)}
    }

    \startdata
00011217+1535355 & 00:01:12 & +15:35:36 & 15.522 ± 0.061 & ... & L4 & MKO $J$ & 3 & 30 & 5.7 \\ 
00144919-0838207 & 00:14:50 & $-08$:38:22 & 14.469 ± 0.026 & L0 & M9 & MKO $J$ & 2 & 20, 30 & 4.6, 5.3 \\ 
00191165+0030176 & 00:19:12 & +00:30:18 & 14.921 ± 0.035 & L1 & L0.5 & MKO $J$ & 2 & 20, 60 & 6.3, 4.0 \\ 
00242463-0158201 & 00:24:25 & $-01$:58:17 & 11.992 ± 0.033 & M9.5 & L0.5 & MKO $J$ & 2 & 10 & 1.1 \\ 
00261147-0943406 & 00:26:11 & $-09$:43:40 & 15.601 ± 0.067 & L1 & ... & MKO $J$ & 2 & 30, 60 & 10.0, 9.1 \\ 
00302476+2244492 & 00:30:25 & +22:44:47 & 14.586 ± 0.036 & ... & L0.5 & MKO $J$ & 3 & 30 & 2.8 \\ 
00304384+3139321 & 00:30:44 & +31:39:31 & 15.480 ± 0.052 & L2 & L3 & MKO $J$ & 3 & 60 & 3.6 \\ 
00320509+0219017 & 00:32:05 & +02:19:01 & 14.324 ± 0.023 & L1.5 & M9 & MKO $J$ & 2 & 30 & 2.9 \\ 
00345684-0706013 & 00:34:57 & $-07$:06:00 & 15.531 ± 0.059 & L3 & L4.5 & MKO $J$ & 2 & 30, 60 & 9.5, 7.1 \\ 
00361617+1821104 & 00:36:16 & +18:21:10 & 12.466 ± 0.025 & L3.5 & L4 & MKO $J$ & 3 & 30 & 1.4 \\ 
    \enddata
    \tablecomments{A standard deviation is listed for data taken with each unique exposure time. Coordinates, $m_J$ values, and spectral types are sourced from the List of Ultracool Dwarfs (\url{https://jgagneastro.com/list-of-ultracool-dwarfs/}).}

\end{deluxetable*}

\section{PINES Transit Search Algorithm} \label{sec:transit_search_and_recovery_algorithm}
PINES data present unique challenges for transit detection. We perform our observations on groups of four to seven L and T dwarf targets which we repeatedly cycle over once per hour throughout observing nights. This is done to maximize the survey efficiency with a single telescope. During a cycle, each target is observed for an amount of time that we refer to as a block, with the block duration determined such that a full cycle of the targets is completed once per hour. For example, in a group of five targets, the block length is set to 10 minutes, with 2 minutes allowed for slewing between targets. The one-hour cycle length was found by \citet{Tamburo2019} to provide an optimal balance between minimizing photon noise and maximizing the likelihood of observing the short-duration ($\sim$50-minute) transits expected for L and T dwarfs. Each group is typically scheduled for five nights of observations at a time, which was found by \citet{Tamburo2019} to maximize the number of expected transit detections in the survey, assuming that planets around L and T dwarfs have the same period-radius distribution as those around M dwarfs \citep{Dressing2015}. 
 
Our observing cadence heavily biases us to the detection of single, incomplete transit events. This bias renders traditional transit search algorithms that rely on the occurrence of multiple transits, like Box Least-Squares \citep{Kovacs2002} and Transit Least-Squares \citep{Hippke2019}, ineffective. Transit identification is further complicated by the fact that L and T dwarfs are frequently observed to be variable on hours-long timescales due to rotation and the presence of heterogeneous surface features \citep[e.g.,][]{Radigan2014, Metchev2015, Vos2019, Vos2020, Tannock2021}. Within the L/T transition (spectral types $\sim$L9$\textendash$T3.5), a ground-based J-band survey established that $39^{+16}_{-14}\%$ of objects exhibit peak-to-peak variability amplitudes greater than 2\% \citep{Radigan2014}. In the L spectral class as a whole, \citet{Metchev2015} found that $80^{+20}_{-27}\%$ of objects are variable with amplitudes greater than $0.2\%$, while $36^{+26}_{-17}\%$ of T dwarfs vary with amplitudes greater than $0.4\%$. However, we note that these observations were performed with the Spitzer Space Telescope in the mid-IR, and the range of amplitudes observed at these wavelengths are generally lower than in the NIR.

An extra degree of caution must therefore be applied when searching for transits around L and T dwarfs, as what appears to be a transit signature may be the result of host variability that is unknown beforehand. When searching for such transits, we require a model that can flexibly account for this potential photometric variability (which may evolve over time) without any prior knowledge about the physical processes occurring on the target's surface. 

A Gaussian process (GP) can serve as such a model. GPs can be used to infer the set of functions that are consistent with time series data as subject to the rules of the chosen covariance function (or \textit{kernel}), which describes the degree of correlation between pairs of data points \citep[for an overview of GPs, see][]{Rasmussen2006}. Quasi-periodic (QP) kernels, in particular, have been used extensively in the literature to search for rotationally induced variability in stars \citep{Pont2013, Haywood2014, ForemanMackey2017, Angus2018, Luque2019} and brown dwarfs \citep{Littlefair}. With the \texttt{celerite} implementation of GP models \citep[][]{ForemanMackey2017}, the QP kernel takes the form

\begin{equation}
    \label{eq:qp_gp_kernel}
    k(\tau) = \frac{B}{2+C}e^{-\tau/L}\left[\cos{\left(\frac{2\pi\tau}{P_{rot}}\right)} + (1+C)\right].
\end{equation}
Here, $\tau$ is a matrix constructed from the array of times $t$, with $\tau_{ij} \equiv |t_i-t_j|$. $B$, $C$, $L$, and $P_{rot}$ are the kernel's hyperparameters, with $B$ and $C$ being related to the amplitude and phase shift of the cosine wave, $L$ representing the timescale over which variability dies off, and $P_{rot}$ representing the rotation period of the target object. 

We used quasi-periodic Gaussian process (hereafter QP-GP) models to develop a transit detection algorithm that addresses the challenges presented by our observing strategy and potential source variability. The algorithm consists of the following steps:

\begin{figure*}[!t]
    \centering
    \includegraphics[width=\textwidth]{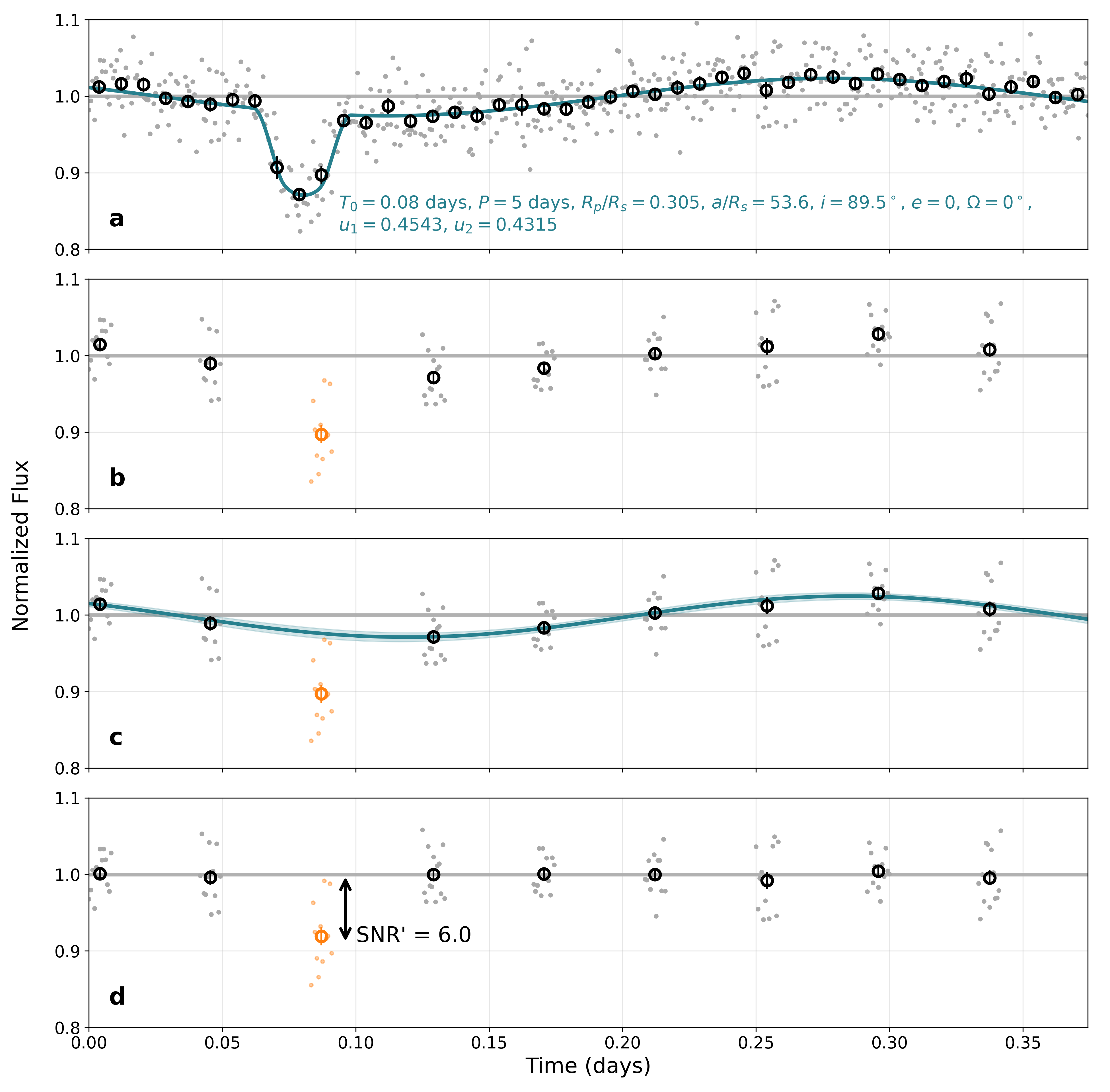}
    
    \caption{An example of a transit recovery in a single night of synthetic data. Panel \textit{a} shows a simulated light curve of a time-variable target. The grey points show photometry with a 60-s exposure time and the black circles with error bars show the data binned over 12 minute intervals. We injected a transiting planet into the light curve with the properties given in the lower-right of panel \textit{a}. The blue line shows the composite transit + variability model. In panel \textit{b}, the light curve is re-sampled to mimic PINES observations, with 12-minute blocks of data obtained once per hour. The third block of data (colored in orange) is decremented by more than 2$\sigma$ and initiates the QP-GP model fit shown in panel \textit{c}. The blue curve in panel \textit{c} shows the mean of the QP-GP model that was fit to the un-binned data excluding points in the third block, and the shaded region indicates the 1$\sigma$ uncertainty of the model. In panel \textit{d}, the flux has been corrected for the QP-GP model. The corrected SNR of the third block is indicated with an black arrow that is offset for clarity.}
    
    \label{fig:simulated_transit_recovery}
\end{figure*}

 \begin{enumerate}
    \item We break an object's single-night light curve into $N$ blocks, the boundaries of which are determined by discontinuities introduced by the PINES cycling strategy. The algorithm operates on single nights of data because the variability signatures of L and T dwarfs can evolve significantly from night to night \citep[e.g.,][]{Artigau2009, Gillon2013b}, an effect observed in PINES light curves themselves \citep{Tamburo2022}. We calculate the mean flux $\bar{F}$ and the error on the mean $\sigma_{\bar{F}}$ of each block\footnote{To be clear, $\sigma_{\bar{F}}$ is the standard deviation of the data in the block divided by the square root of the number of points in the block.}. 
    
    \item We loop over the $N$ blocks. At block $i$ of $N$, we calculate the mean $\mu$ and standard deviation $\sigma$ of the binned data excluding the $i^{th}$ block:
    
    \begin{equation}
        \mu = \frac{1}{N-1}\sum_{j=1, j \neq i}^{N} \bar{F_j};
    \end{equation}
    
    \begin{equation}
        \sigma = \sqrt{\frac{\sum_{j=1, j \neq i}^{N} (\bar{F_j}-\mu)^2}{N-1}}.
    \end{equation}
    
    \item We calculate the signal-to-noise ratio (SNR) of the difference between $\mu$ and $\bar{F_i}$:
    
    \begin{equation}
        SNR = \frac{\mu-\bar{F}_i}{\sqrt{\sigma^2+\sigma_{\bar{F}_i}^2 }}
        \label{eq:snr}
    \end{equation}

    \item A positive SNR indicates that the block in question is decremented compared to the mean of the rest of the binned data. For efficiency, we continue our search only if $SNR > 2$, using \texttt{celerite} to initialize a QP-GP model with the kernel given in Equation \ref{eq:qp_gp_kernel} and a mean of 1. We use the L-BFGS-B nonlinear optimization routine \citep{Byrd1995, Zhu1997} as implemented in the \texttt{scipy} Python package \citep{SciPy-NMeth2020} to estimate the parameter values that best fit the light curve excluding the $i^{th}$ block, subject to the bounds in Table \ref{tab:gp_limits}. The bounds on $B$ and $C$ permit peak-to-peak variability amplitudes up to $\sim$26\%, which is consistent with the largest measured amplitude for a brown dwarf \citep[2MASS J21392676+0220226;][]{Radigan2012}. The bounds on $L$ allow for exponential decay timescales up to eleven days, the maximum extent of PINES light curves. Finally, the bounds on $P_{rot}$ permit rotation periods down to a limit of two hours, below which PINES observations are generally not Nyquist sampled due to our cycling strategy. 
    
    With four free parameters fit to N-1 blocks, we require the night being searched to consist of at least five blocks of data (i.e., a time extent of about five hours or more). This is true of all nights in the light curve sample described in Section \ref{sec:observations}.
    
\begin{deluxetable}{lrr}
    \tablecaption{Lower and upper bounds on the QP-GP hyperparameters used in our transit search algorithm.}
    \tablenum{2}
    \label{tab:gp_limits}

    \tablehead{
        \colhead{Parameter} & \colhead{Lower Bound} & \colhead{Upper Bound}
        }
        
    \startdata
        $\ln{B}$ &  -20 & -1.3\\
        $\ln{C}$ & -20 & -1.3\\
        $\ln{L}$ & -20 & 2.4 \\
        $\ln{P_{rot}}$ & -2.5 & 3.5 \\
    \enddata
    
\end{deluxetable}

\item We divide the flux by the optimized QP-GP model. We recalculate $\mu$, $\sigma$, $\bar{F_i}$, and $\sigma_{\bar{F_i}}$ using the corrected data, propagating the uncertainty of the QP-GP model into our estimate of $\sigma_{\bar{F_i}}$. Finally, we calculate the corrected SNR of the block in question, SNR$'$, following Equation \ref{eq:snr} but replacing the quantities with their values from the corrected light curve. 

For clarity, we emphasize that SNR values represent the significance with which a given block of data is decremented compared to the mean of the rest of the light curve on a single night. SNR$'$ values, on the other hand, represent the significance with which a block of data is decremented after the light curve has been corrected with a QP-GP model (which has been fit excluding the block in question). The SNR and SNR$'$ values of a particular block will not necessarily be the same because of the QP-GP model correction.

\end{enumerate}

Figure \ref{fig:simulated_transit_recovery} shows an example of the transit search algorithm in action on simulated data. For clarity, we first generated time stamps for nine hours of staring data for a target with 60 s exposure time (panel $a$). We simulated data assuming Gaussian noise with a standard deviation $\sigma = 0.03$, a typical value for an $m_J \approx 14$ PINES target. We then injected a variability signal, with a peak-to-peak amplitude of 5\% and a period of 8 hours. We added a transiting planet model with a planet-to-host radius ratio of $0.305$, which corresponds to a planet with radius $3.4$ $R_\oplus$ around a 1 $R_{Jup}$ brown dwarf. In panel $b$, we re-sampled the staring light curve to mimic a typical PINES data set, with 12-minute blocks of data obtained once per hour. We show the results of our transit search algorithm in panel $c$. The third block is significantly decremented in the corrected light curve with an SNR$'$ = 6.0, and the QP-GP variability model accurately captures the rotational variability signal in the other blocks.


\section{Transit Search} \label{sec:transit_search}
\begin{figure}
    \centering
    \includegraphics[width=\columnwidth]{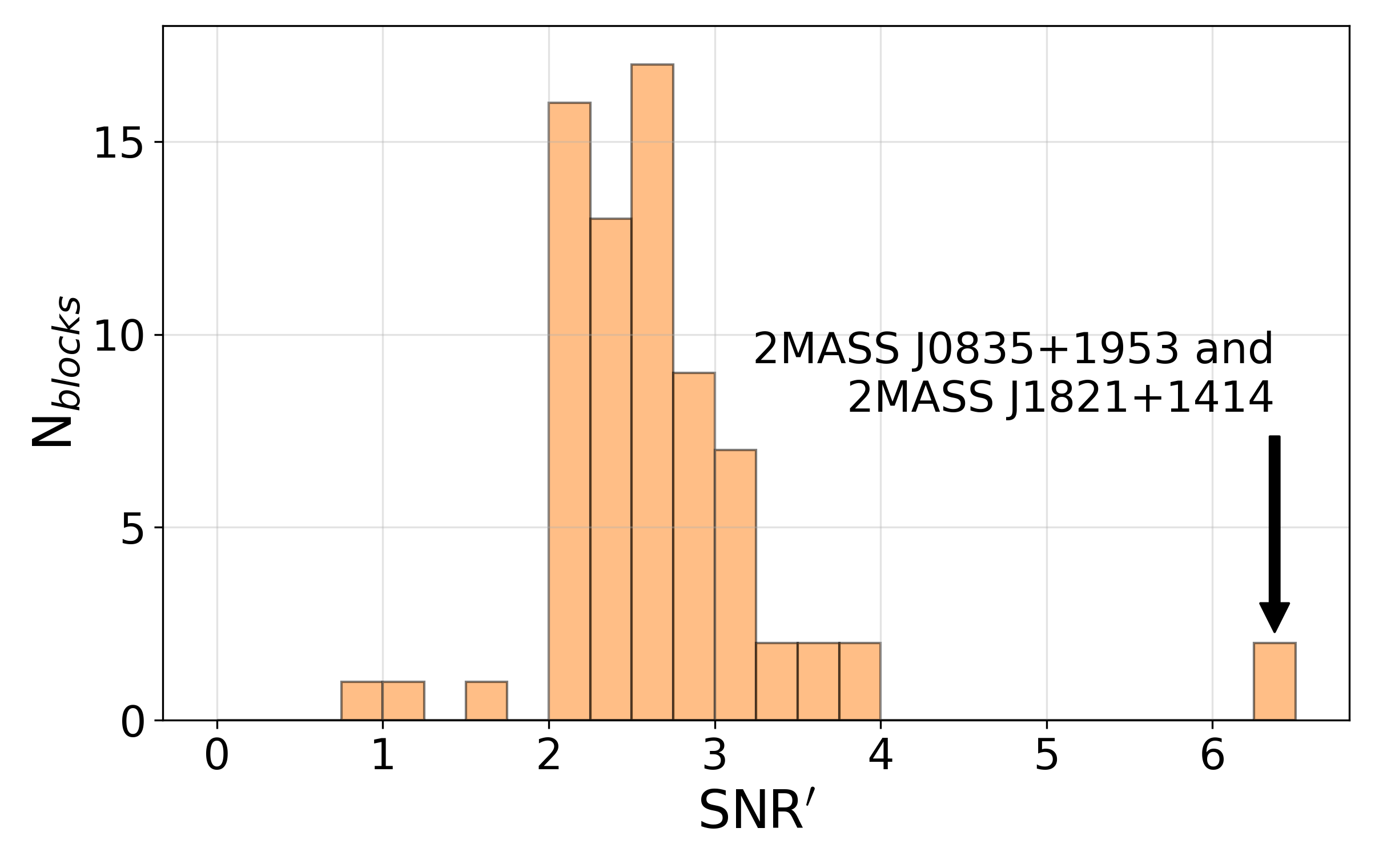}
    \caption{The distribution of SNR$'$ values for the 73 blocks of our light curve sample that are decremented to more than 2$\sigma$.}
    \label{fig:snr_distribution}
\end{figure}

We used our algorithm to search for transits in the light curve sample described in Section \ref{sec:observations}. These light curves consist of 2999 individual blocks of data, 73 of which were found to be decremented with an SNR greater than 2. This is close to the number expected for a Gaussian distribution consisting of 2999 samples ($\sim68$). We calculated the SNR$'$ values for these 73 blocks following steps 4 and 5 from the transit detection algorithm described in the previous section. We show the distribution of the SNR$'$ values in Figure \ref{fig:snr_distribution}. Two of the blocks were much more significantly decremented than the rest of the sample, with SNR$'$ values greater than six. We discuss these two candidate events in detail in the following sections. 

\subsection{2MASS J18212815+1414010}
\label{sec:2m1821}

\begin{figure*}[t]
    \centering
    \includegraphics[width=\textwidth]{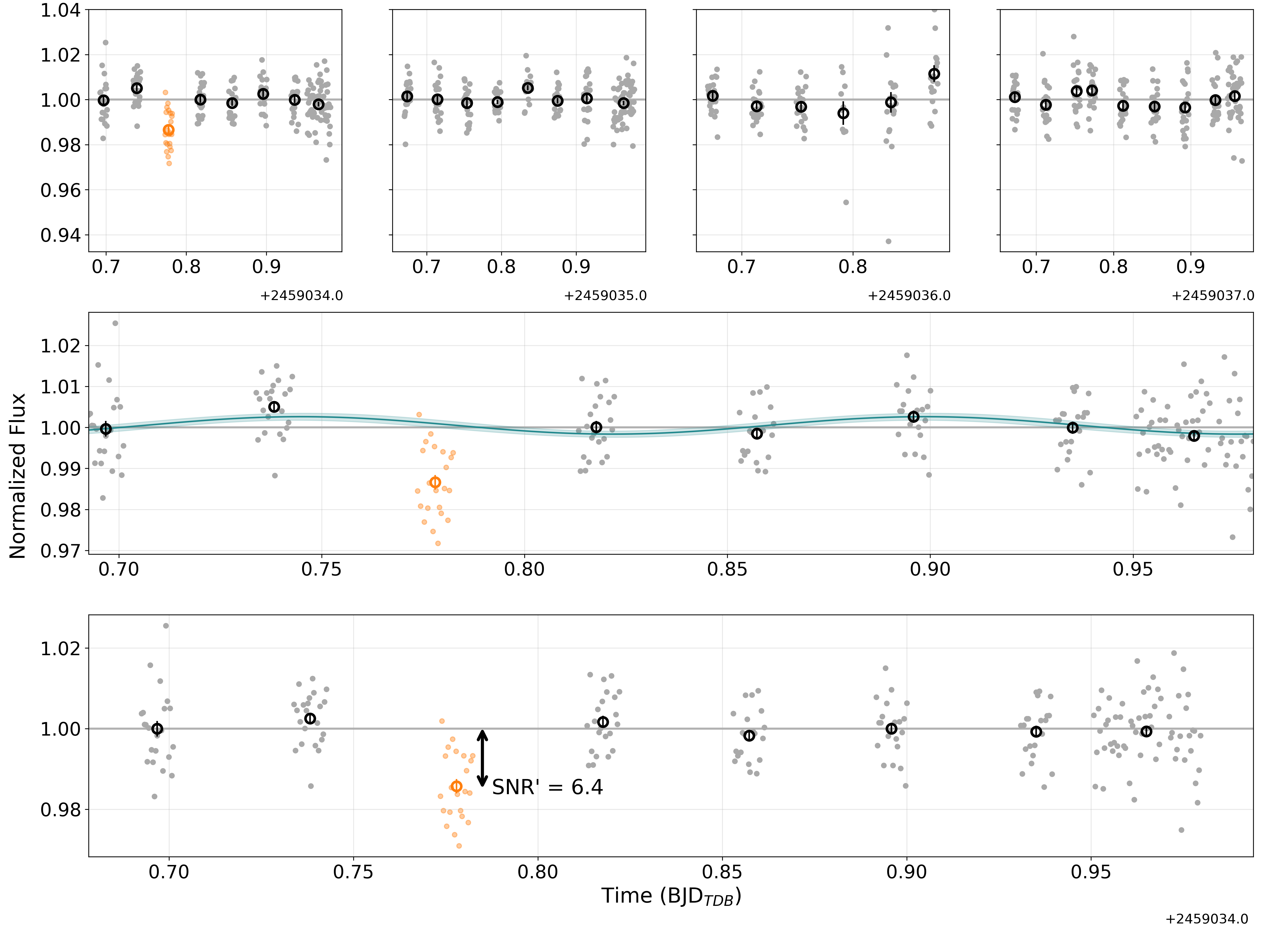}
    \caption{Top: the full PINES light curve for 2MASS J1821+1414. Grey points show the 30 s photometry and black points with error bars show the photometry binned over the duration of each block. The alleged transit event (the third block on the first night) is colored in orange. The light curve was normalized ignoring this block of data. Middle: The first night of the 2MASS J1821+1414 light curve.} The blue line shows mean of the QP-GP model fit using unbinned data, excluding the block of the candidate transit. The blue shaded area indicates the 1$\sigma$ uncertainty on the model. Bottom: The light curve corrected for the QP-GP model shown in the middle panel. The SNR$'$ of the suspected transit block, calculated as described in Section \ref{sec:transit_search_and_recovery_algorithm}, is indicated with an offset arrow.
    \label{fig:2m1821}
\end{figure*}

2MASS J18212815+1414010 (2MASS J1821+1414) is a brown dwarf that was discovered in a proper-motion search of 2MASS data, having gone previously undetected due to its peculiarly red NIR colors ($J-K_S = 1.78 \pm 0.05$) and its location in a crowded field within $15^\circ$ of the Galactic plane \citep{Looper2008}. It was classified with an optical spectral type of L4.5 and an NIR spectral type of L5 pec \citep[denoting the aforementioned peculiarly red NIR spectrum,][]{Looper2008}. It is an unlikely member of any known young moving groups \citep{Gagne2014}. It has a mass estimate of $0.049^{+0.014}_{-0.024}$ $M_\odot$ from evolutionary models \citep{Chabrier2000}, an age estimate of $120-700$ Myr, and is located at a distance of $9.38 \pm 0.03$ pc \citep{Sahlmann2016}. Being nearby, 2MASS J1821+1414 is relatively bright, with an apparent 2MASS \textit{J}-band magnitude of $13.4$ (placing it in the 91\textsuperscript{st} percentile of $J$-band magnitudes in the PINES sample). 

We obtained four nights of PINES observations of 2MASS J1821+1414 from UT 2020 July 04 to 2020 July 07 in J-band with a $30$ s exposure time. Observing conditions were generally clear, with some clouds on the third night. The average seeing FWHM of the observations was $2.^{\prime\prime}3 \pm 0.^{\prime\prime}5$ and we found that time-variable apertures with radii set to $1\times$ a smoothed trend of the seeing provided the highest precision light curve. The target's raw flux was corrected using an ALC made up of the weighted flux of 28 reference stars \citep[for details on the ALC creation procedure, see][]{Tamburo2022}. These reference stars bracket the target in average brightness, ranging from $0.6-3.0$ times its average flux. We show the full PINES light curve for 2MASS J1821+1414 in the top row of Figure \ref{fig:2m1821}. 

Our transit search algorithm identified a candidate event on the first night, which we show in greater detail in the bottom panel of Figure \ref{fig:2m1821}. The algorithm found that the third block of data on this night was decremented below a QP-GP model optimized on the other blocks with an SNR of 6.4. The candidate event has a depth of about $1.4$\%, and the QP-GP model that was used to identify it has a peak-to-peak amplitude of $0.4\pm0.2$\% with a rotation period of $3.8\pm0.5$ hr. 


However, 2MASS J1821+1414 is a known variable and was first identified as such in Spitzer Space Telescope/Infrared Array Camera time series observations. \citet{Metchev2015} reported peak-to-peak variability amplitudes of $0.54 \pm 0.05$\% at 3.6 $\mu$m and  $0.71 \pm 0.14$\% at 4.5 $\mu$m, noting that the variability was irregular with a best-fit period of $4.2 \pm 0.1$ hr. The authors suggested that the light curve's time evolution and multi-peaked periodogram could be attributed to rapid changes in the distribution of clouds across the photosphere, possibly through multiple cloud layers. This interpretation was bolstered by observations of a $\sim$180$^\circ$ phase shift between simultaneous Spitzer and Hubble Space Telescope (HST)/Wide Field Camera 3 light curves, which probe different depths in the atmosphere \citep{Yang2016}. Percent-level aperiodic variability has also been established in $i$-band, with different peak-to-peak amplitudes reported for different epochs \citep{MilesPaez2017}. Finally, \citet{Schlawin2017} performed time-resolved Infrared Telescope Facility (IRTF)/SpeX observations of 2MASS J1821+1414 over a wavelength range of $0.9–2.4$ $\mu$m, finding $\sim2$\% peak-to-peak variability amplitudes across J-band. Given the known variability properties of 2MASS J1821+1414, we disfavor a transiting planet hypothesis for explaining its PINES light curve. The object is a known irregular variable and has been observed to exhibit percent-level flux changes in $J$-band. 

\subsection{2MASS J08350622+1953050}
\label{sec:2m0835}

\begin{figure*}[ht]
    \centering
    \includegraphics[width=\textwidth]{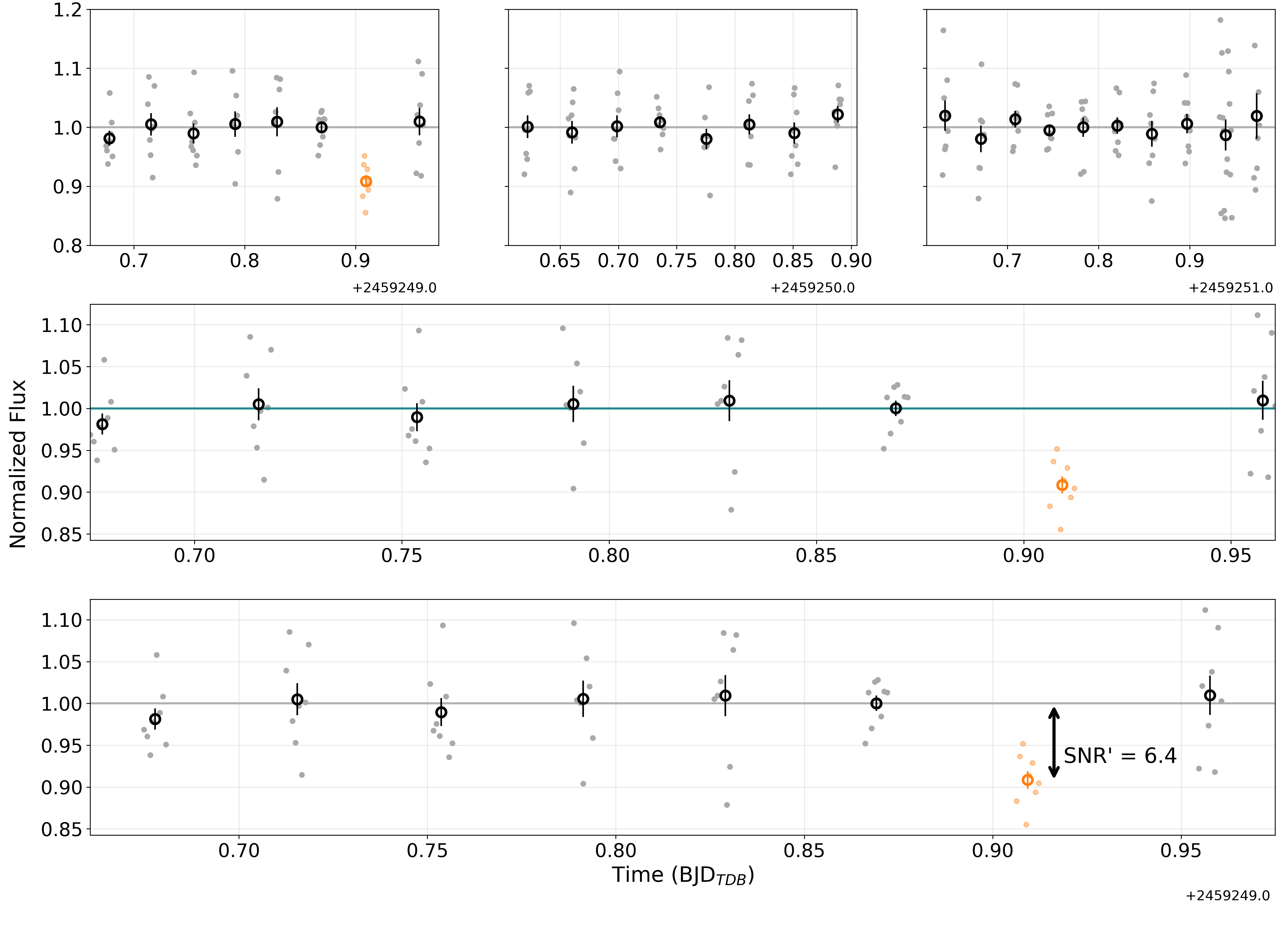}
    \caption{Top: The full PINES light curve for 2MASS J0835+1953. Middle: The first night of data, showing the candidate transit event in more detail. The plot elements match those of Figure \ref{fig:2m1821}. The optimized QP-GP model fit no apparent variability and appears as a flat line with a narrow 1$\sigma$ confidence region. Bottom: The corrected light curve, which is largely unchanged due to the flat QP-GP model in the middle panel. The SNR$'$ of the suspected transit event is indicated with an offset arrow.}
    \label{fig:2m0835}
\end{figure*}

Decidedly less is known about our second transit candidate, 2MASS J08350622+1953050 (2MASS J0835+1953), likely owing to its relative faintness (2MASS $m_J = 16.1$, 2\textsuperscript{nd} percentile of $J$-band magnitudes in the PINES sample). 2MASS J0835+1953 was identified in a search for L and T dwarfs in Sloan Digital Sky Survey (SDSS) data with an $i-z$ color of $2.26 \pm 0.09~$mag and classified with an NIR spectral type of L4.5 with IRTF/SpeX spectroscopy \citep{Chiu2006}. \citet{Kirkpatrick2010} revised its NIR spectral type slightly to L5 and identified it as an NIR spectral standard. It has a surface gravity that is consistent with that of a field object \citep{Martin2017} and is located at a distance of $36 \pm 4$ pc \citep{Faherty2009}. No time series photometry has been published for this object.  

We show the full PINES light curve for 2MASS J0835+1953 in the top row of Figure \ref{fig:2m0835}. We observed the target for three nights, from UT 2021 February 4 to 2021 February 6, using 60 s exposures in \textit{J}-band. Observing conditions were mostly clear with some light cirrus clouds on the first and third nights, and an average seeing FWHM of $2.^{\prime\prime}8 \pm 0.^{\prime\prime}6$. The target's ALC was created using flux measurements from 23 reference stars. The reference stars bracket 2MASS J0835+1953 in average brightness, ranging from $0.8-2.8$ times its average flux. The optimum light curve used time-variable apertures with a radii set to $1\times$ a smoothed profile of the seeing FWHM. 

Our transit detection algorithm identified a candidate event during the first night with a depth of $8.5\pm1.2\%$ and an SNR of 6.4, which is shown in more detail in the bottom panel of Figure \ref{fig:2m0835}. The QP-GP model finds no evidence for significant variability in the light curve. 

\subsubsection{Diagnostic Checks}
\label{sec:diagnostic_checks}
Unlike 2MASS J1821+1414, we are unable to rule out the planetary nature of this candidate event on the basis of source variability. Here, we perform a series of diagnostic checks to see if 2MASS J0835+1953's flux dimming event can be explained with systematic effects. 

One potential source of systematic flux variation is changes in the field positioning coupled with flat fielding errors. PINES observations are performed using a custom guiding procedure that corrects image shifts with a fast Fourier transform between science images and a pre-defined ``guide field" image \citep{Tamburo2022}. This guiding approach can apply inaccurate shifts in images with poor guiding solutions (due to, e.g., clouds or a source detection FWHM that is not close to the true seeing FWHM).

\begin{figure}
    \centering
    \includegraphics[width=\columnwidth]{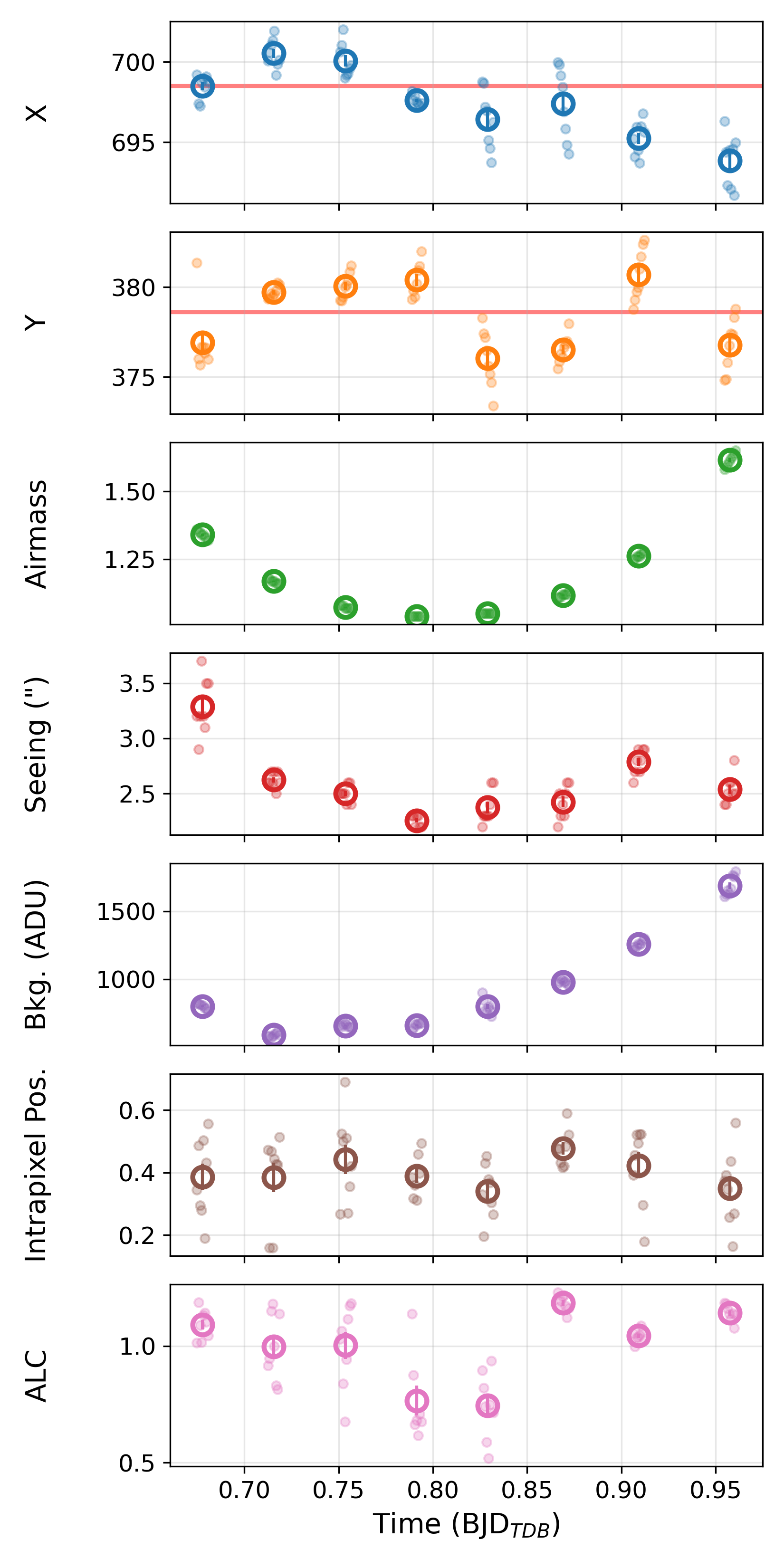}
    \caption{Diagnostic time series on the night of 2MASS J0835+1953's candidate transit. From top to bottom: $x$ centroid position, $y$ centroid position, airmass, seeing FWHM ($^{\prime\prime}$), sky background (ADU), intrapixel position, and ALC. Unfilled circles with error bars show the mean of individual measurements (filled circles) over each block. In the $x$ and $y$ centroid panels, red lines indicate the intended guide field position of the target. None of the diagnostics were found to correlate significantly with target flux.}
    \label{fig:diagnostics}
\end{figure}

We plot the target's measured $x$ and $y$ centroid positions on the night of the candidate event in the top two panels of Figure \ref{fig:diagnostics}. While there was a general leftward drift in the target's $x$ pixel position, we found no evidence for anomalously large image shifts. The maximum distance between a measured field position and the guide field position (indicated with red lines in the top two panels of Figure \ref{fig:diagnostics}) was $6.8$ pixels, with a mean distance of $3.2 \pm 1.5$ pixels. These values are consistent with our typical guiding performance, leading us to conclude that inaccurate field placement was not the cause of the candidate event. We also visually inspected the measured centroids and verified that each was centered on the target, meaning that the transit candidate cannot be attributed to inaccurately placed apertures. 

\begin{figure}
    \centering
    \includegraphics[width=\columnwidth]{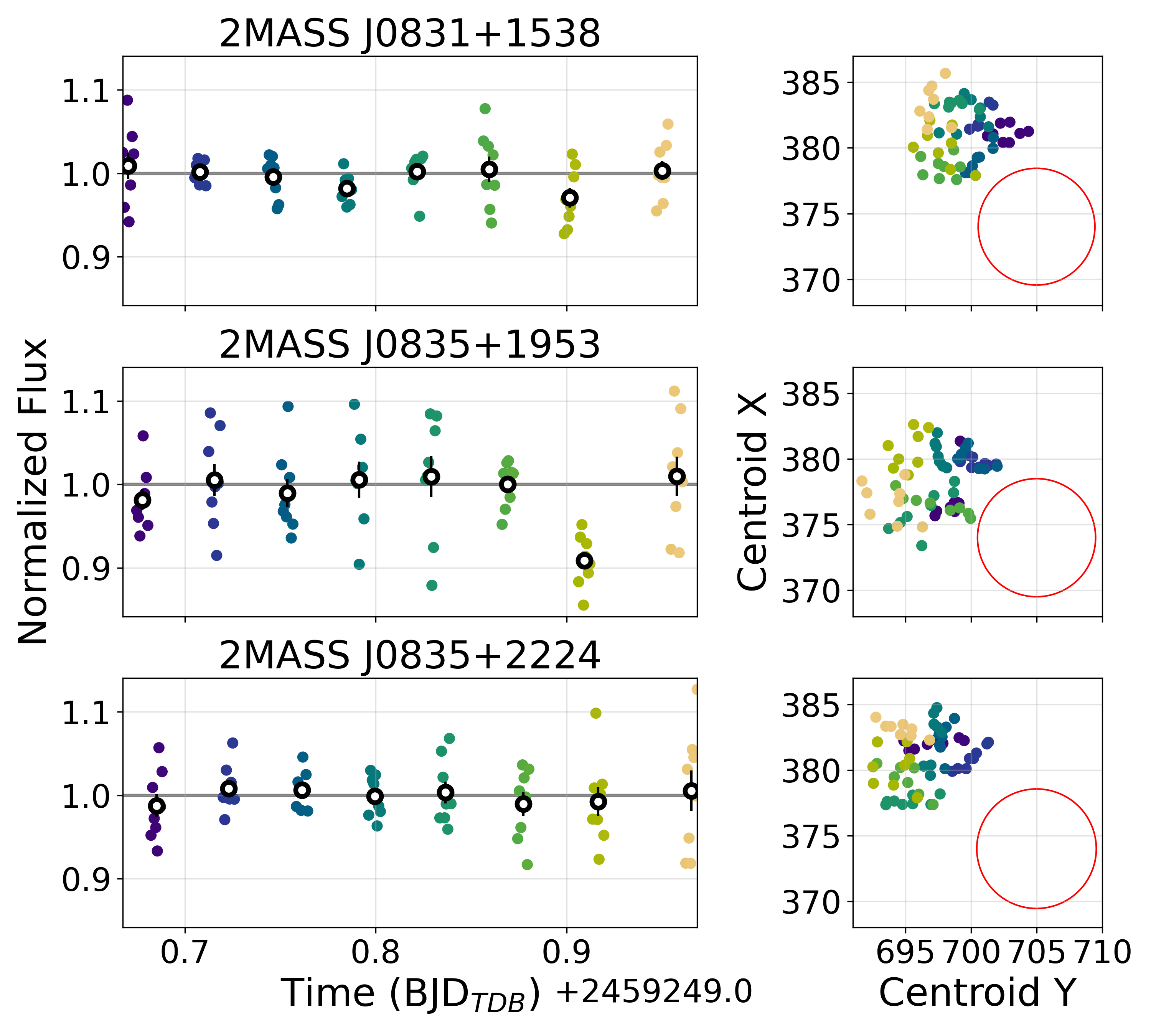}
    \caption{The PINES light curves of 2MASS J0831+1538, 2MASS J0835+1953, and 2MASS J0835+2224 on the night of 2MASS J0835+1953's candidate transit event. All three light curves have been normalized ignoring the seventh block of data. Unbinned points are color-coded by time. The right-hand column shows the measured centroid positions, and the red circles show the average aperture size used for performing variable aperture photometry. For ease of comparison, the plot limits are the same in each column.}
    \label{fig:lc_comparison}
\end{figure}

As a second check on flat fielding errors, we examined the light curves of the targets observed just before and just after 2MASS J0835+1953 in the observing group. Our targets are deliberately placed on the same detector position in order to avoid bad pixels \citep{Tamburo2022}, so they roughly sample the same pixels over the course of the night. Signals introduced by flat fielding errors are therefore likely to manifest themselves in the light curves of objects observed nearby in time. We show the light curves of 2MASS J0831+1538, 2MASS J0835+1953, and 2MASS J0835+2224, which were observed consecutively on the night of the candidate transit event, in the left-hand column of Figure \ref{fig:lc_comparison}. The unbinned data in these light curves are color coded by their time stamp. In the right-hand column, we show the measured centroid positions of the targets, with the colors matching those in the left-hand column. We also show the average aperture size that was used to perform the variable aperture photometry as a red circle.

Of particular importance for this analysis is the relative positions of the three targets during the seventh block, the alleged in-transit point for 2MASS J0835+1953: a comparison of this sort would not be valid if 2MASS J0835+1953 was placed on substantially different pixels during than the other two targets. During the seventh block, the difference between the mean $x$ and $y$ centroid positions for 2MASS J0831+1538 and 2MASS J0835+1953 was $2.6\pm1.7$ and $0.6\pm2.0$ pixels, respectively. For 2MASS J0385+1953 and 2MASS J0835+2224, these values were $1.0\pm1.6$ and $0.5\pm2.0$ pixels. These small differences imply that 2MASS J0831+1538 and 2MASS J0835+2224 sampled roughly the same pixels as 2MASS J0835+1953 during the candidate transit block, and therefore their light curves should be suitable for testing for flat fielding errors.

The seventh block of 2MASS J0831+1538's light curve shows a slight decrement compared to the rest of its binned data, with a significance of $2.2\sigma$. The same block of the 2MASS J0835+2224 light curve is decremented by $0.4\sigma$. Because neither light curve shows a significant flux decrease during the seventh block, and because both sample roughly the same pixels as 2MASS J0835+1953, we conclude that the candidate transit event cannot be attributed to flat fielding errors.

Another potential systematic source of a flux decrement in PINES data is the non-linearity of the Mimir detector. Through testing, we have determined that the detector provides a linear response to within uncertainties below $\sim$4000 ADU, but that measured counts are systematically underestimated above this threshold \cite[consistent with linearity characterization tests of the Mimir detector presented in][]{Clemens2007}. This effect grows to about $10\%$ by $6500$ ADU. We examined the peak source counts of 2MASS J0835+1953 and its 23 reference stars within the photometric apertures in all images on the night of the candidate event, and found a maximum single pixel value of $3629$ ADU. This is well below the level needed to impart an $8.5\%$ transit signal, and we therefore cannot attribute the candidate event to non-linearity effects.

An inaccurate modeling of bad pixel values could also lead to a flux dimming in PINES data. The Mimir detector contains a large number of dead, hot, and variable pixels (about 2.7\% of the detector, on average), and if these pixels fall within a source aperture, we model them with a 2D Gaussian replacement procedure \citep{Tamburo2022}. If the target encounters a bad pixel and its point spread function deviates significantly from a 2D Gaussian, our approach could feasibly underestimate the true target flux, leading to a flux decrease. However, our pipeline records whether or not a bad pixel correction is needed within a source aperture in every image, and we found that 2MASS J0835+1953 required no such corrections on the night of the candidate event. We can thus rule out an inaccurate bad pixel correction as a potential cause.

Alternatively, a flux dimming of several percent could be introduced by the failure to flag a time-variable bad pixel near the center of the target's PSF. Specifically, with $2".8$ FWHM seeing and $0".579$ pixels, a transient bad pixel at the center of the PSF could cause a flux decrease as large as $\sim$5\% if its quantum efficiency dropped entirely to zero during the candidate event. However, we find this scenario to be an unlikely explanation of the observed flux dimming. First, the target was not held on precisely the same pixel position throughout the candidate event, with a mean $x$ centroid of $695.2\pm1.0$ and a mean $y$ centroid of $380.7\pm1.4$ (see Figures \ref{fig:diagnostics} and \ref{fig:lc_comparison}). An unidentified transient bad pixel would therefore not coincide with the peak of the target's PSF throughout the candidate event and could not impart a dimming effect on the order of 5\%. Second, we inspected the images during the candidate event, and found no evidence of a transient dark pixel that developed within the target's PSF.

The target flux is corrected with an ALC, which is a weighted sum of normalized reference star fluxes (shown in the bottom panel of Figure \ref{fig:diagnostics}). These weights are calculated in an iterative procedure which is designed to de-weight variable reference stars \citep{Murray2020, Tamburo2022}. In order to account for 2MASS J0835+1953's light curve, the weighting procedure would need to highly weight a reference star (or set of reference stars) that underwent a sudden flux brightening event during the alleged in-transit block, and we find that this is not the case. The ALC creation procedure assigned a maximum reference star weight of $15.5\%$ and a minimum weight of $0.5\%$, and no simultaneous brightening event was seen in the reference star fluxes during the candidate transit event.

After correcting with the ALC, we perform a final linear regression correction on our target light curves. We use the \texttt{pearsonr} function as implemented in \texttt{Scipy} to calculate the significance of correlation between the ALC-corrected target flux and a variety of regressors, including airmass, centroid $x$ and $y$ positions, seeing, sky background, and the target's intrapixel location. We show time series of these regressors on the night of the candidate event in Figure \ref{fig:diagnostics}. Any regressor that has a two-tailed $p$-value less than $0.01$ is retained in the regression, otherwise it is discarded. No regressors were significantly correlated with the ALC-corrected target flux on the first night of 2MASS J0835+1953's light curve, so this step could not have caused the candidate transit event. 

Time-variable PWV can introduce signatures that mimic transit events through \textit{second-order extinction}, in which stars with different spectral energy distributions (SEDs) experience different levels of wavelength-integrated extinction \citep[e.g.,][]{BailerJones2003,Blake2008,Baker2017,Murray2020}. \citet{Tamburo2022} modeled the expected magnitude of this effect in MKO \textit{J}-band, finding that an L6 target will undergo a 0.4\% flux decrement when corrected with a G0 reference star over a PWV range of 10 mm. With this same model, we estimate that a rapid and drastic increase of several thousands of mm of PWV would be required to explain an $8.5\%$ flux dip in MKO \textit{J}-band. The required increase is roughly two to three orders of magnitude higher than typical atmospheric PWV measurements at astronomical sites \citep[e.g.,][]{Cortes2020, Li2020, Murray2020, MeierValdes2021}. Thus, PWV changes are an exceedingly unlikely explanation for the flux decrease in 2MASS J0835+1953's light curve.

Finally, we considered the possibility that effects from high cirrus clouds on the night of the candidate event caused the flux dimming. We note that if clouds introduced systematic flux changes to the final target light curve, their signatures would also likely be present in the light curves of the objects observed just before and after 2MASS J0835+1953 (see Figure \ref{fig:lc_comparison}). No such correlation is seen, but this does not rule out the possibility of a rapid patch of cirrus clouds affecting the seventh block of 2MASS J0835+1953's light curve alone. This scenario, however, can be ruled out by inspecting the ALC, shown in the bottom panel of Figure \ref{fig:diagnostics}. The ALC captures the average flux of sources in the field (excluding the target) throughout the night, and it shows that this average flux was near its highest point of the night during the last three blocks. No significant flux decrease was observed in the ALC during the seventh block, so the flux decrement cannot be attributed to the sudden appearance of clouds.

We also performed an empirical test for a correlation between flux decrements in PINES data and the presence of cirrus clouds. We created a sample of all nights in our light curve sample that had a block with an SNR$'$ value greater than 3, of which there were 15 (see Figure \ref{fig:snr_distribution}). We then looked for evidence of clouds in the raw flux measurements on these nights, finding that only one showed signs of clouds besides the night of 2MASS J0835+1953's candidate event. There is therefore no evidence that light cirrus clouds induce significant flux decrements in PINES data, and we conclude that the candidate event in 2MASS J0835+1953's light curve is not attributable to the presence of cirrus clouds.

As none of the sources of contamination we considered (flat fielding errors, detector non-linearity, bad pixels, reference star weights, regressors, PWV, and clouds) can account for the observed transit event, we turn to astrophysical false positive scenarios as an explanation.

\subsubsection{Astrophysical False Positives}
\label{subsec:false_positives}

A variety of astrophysical scenarios can mimic transit events, generally involving unknown configurations of stellar eclipsing binaries (EBs). For example, the chance alignment of a background EB with a target can dilute the depth of eclipses to a level that resembles a planetary transit \citep[e.g.,][]{Batahla2013}. In a hierarchical triple EB system, a bright star coupled with an EB can achieve the same effect, as was the case for OGLE-TR-33 \citep{Torres2004}. The target itself could be an unknown EB, with grazing eclipses producing transit-like shapes and depths.

\begin{figure*}[!t]
    \centering
    \includegraphics[width=\textwidth]{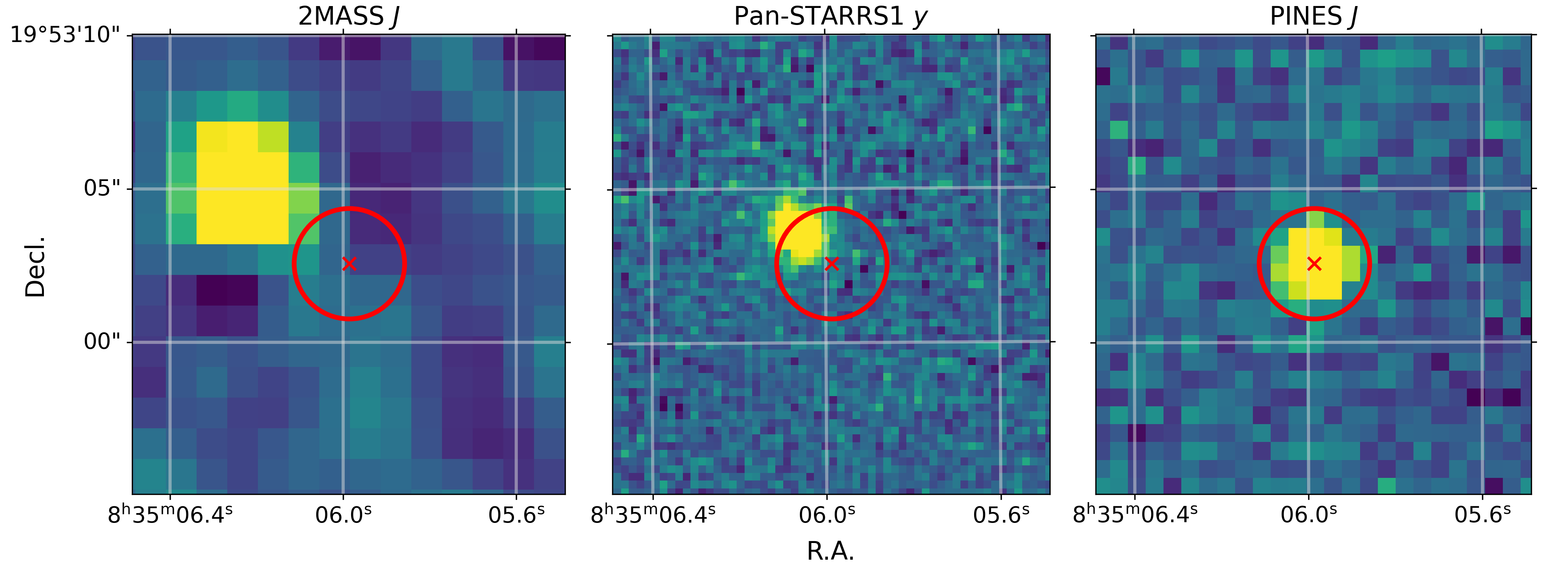}
    \caption{The positions of 2MASS J0835+1953 in 2MASS (Nov. 1998), Pan-STARRS (Apr. 2010\textendash Dec. 2014), and PINES (Feb. 2021) images. All images are centered on the measured position during the PINES epoch. A red x with a surrounding circular aperture indicates the PINES position in all three panels.}
    \label{fig:motion}
\end{figure*}

The background EB scenario can be investigated by taking advantage of the target's high proper motion of $-158\pm13$ mas/yr in R.A. and $-108\pm14$ mas/yr in Decl. \citep{Faherty2009}. We downloaded archival imagery of 2MASS J0835+1953 from 2MASS \citep{2MQuicklook} and Pan-STARRS\footnote{\url{http://ps1images.stsci.edu/cgi-bin/ps1cutouts}} \citep{Kaiser2002}. The 2MASS image was taken in $J$-band Nov. 1998, and the Pan-STARRS image consists of stacked exposures in $y$-band taken from Apr. 2010 to Dec. 2014. We show these images in Figure \ref{fig:motion}, along with a PINES image taken in Feb. 2021. The motion of the target reveals that no bright background sources were present in the PINES epoch, which suggests that the candidate transit event cannot be attributed to a background EB.

\begin{figure}
    \centering
    \includegraphics[width=\columnwidth]{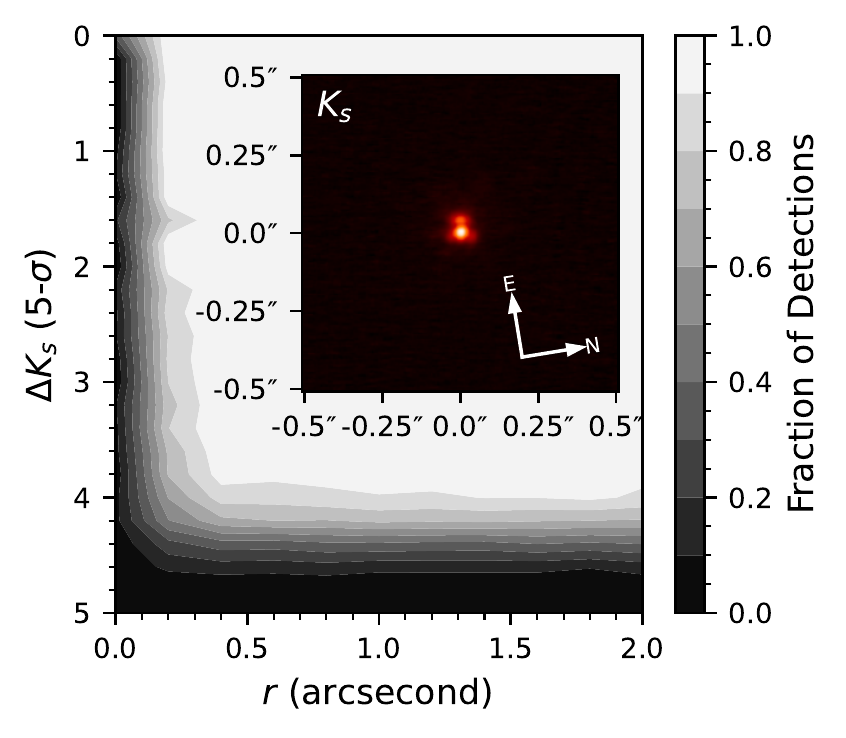}
    \caption{Keck/NIRC2 AO $K_s$-band 5-$\sigma$ contrast curve from our injected planet simulation (Section~\ref{subsec:false_positives}). The inset plot shows the $1\arcsec\times1\arcsec$ cutout of the combined image. No companions are observed within the contrast limits of the observations.}
    \label{fig:ao}
\end{figure}

To investigate the possibility of the source itself being an EB, we obtained adaptive optics (AO) imaging of 2MASS J0835+1953 using the Near-infraRed Camera 2 (NIRC2) and laser guide star (LGS) AO system \citep{vanDam2006, Wizinowich2006} on the Keck II 10-m Telescope on UT 2022 Jan 11. Three K-band images were taken with exposure times of 60s in a box 3 pattern to avoid the noisy bottom left quadrant. Data were reduced using the Keck AO Imaging \citep[KAI;][]{Lu2022} data reduction pipeline, including flat fielding, bias subtraction, dark correction, sub-pixel alignment, and weighting by Strehl. These data have a SNR of 1513, a Strehl ratio of 0.295, and a FWHM of 61.92 mas. The final reduced image is shown in the inset plot of Figure \ref{fig:ao}, with the trefoil pattern clearly visible in the PSF.

To create the 5-$\sigma$ contrast curve, we injected a simulated planet as a Gaussian with a PSF similar to the host brown dwarf, and giving it a random position angle between 0–360$^\circ$, a random distance between 0–2$\arcsec$, and a random magnitude difference between 0–6. Source detections were determined using the IRAF \texttt{starfind} method, requiring a 5-$\sigma$ detection above the background. We repeated this injection test 100,000 times, and show the fraction of detected planets in the contrast curve in Figure~\ref{fig:ao}. There are no detectable companions within the contrast limits of the observations.

We also searched 2MASS J0835+1953's SpeX Prism spectrum\footnote{The spectrum was accessed from \url{https://cass.ucsd.edu/~ajb/browndwarfs/spexprism/html/published.html}.} for evidence of binarity, following the spectral template fitting approach described in \citet{Burgasser2010} and \citet{BardalezGagliuffi2014}. We used 715 template spectra from 622 sources (excluding known and candidate binary systems, known and candidate variable objects, and subdwarfs) to generate a sample of $110,110$ binary templates by randomly adding two flux-normalized spectra together. We compared 2MASS J0835+1953's spectrum to each single (715) and binary ($110,110$) spectral template, calculated the $\chi^2$ statistic for each case, and ranked the single fits and binary fits by $\chi^2$. We then did hypothesis testing with an F-test on the reduced $\chi^2$ values for the best single and binary fits in order to reject or accept the null hypothesis (i.e., that the candidate is not a binary). As a result of the F-tests, we calculate a 75\% confidence that we can reject the null hypothesis, suggesting that there is marginal evidence that 2MASS J0835+1953 is a spectral binary. The best-fit single template was an L4. The best-fit binary template was an L5.5 + T0.0, though the secondary spectral type is uncertain by 2.4 subtypes.

\begin{figure*}[!tbp]
  \centering
  \begin{minipage}[b]{0.49\textwidth}
    \includegraphics[width=\textwidth]{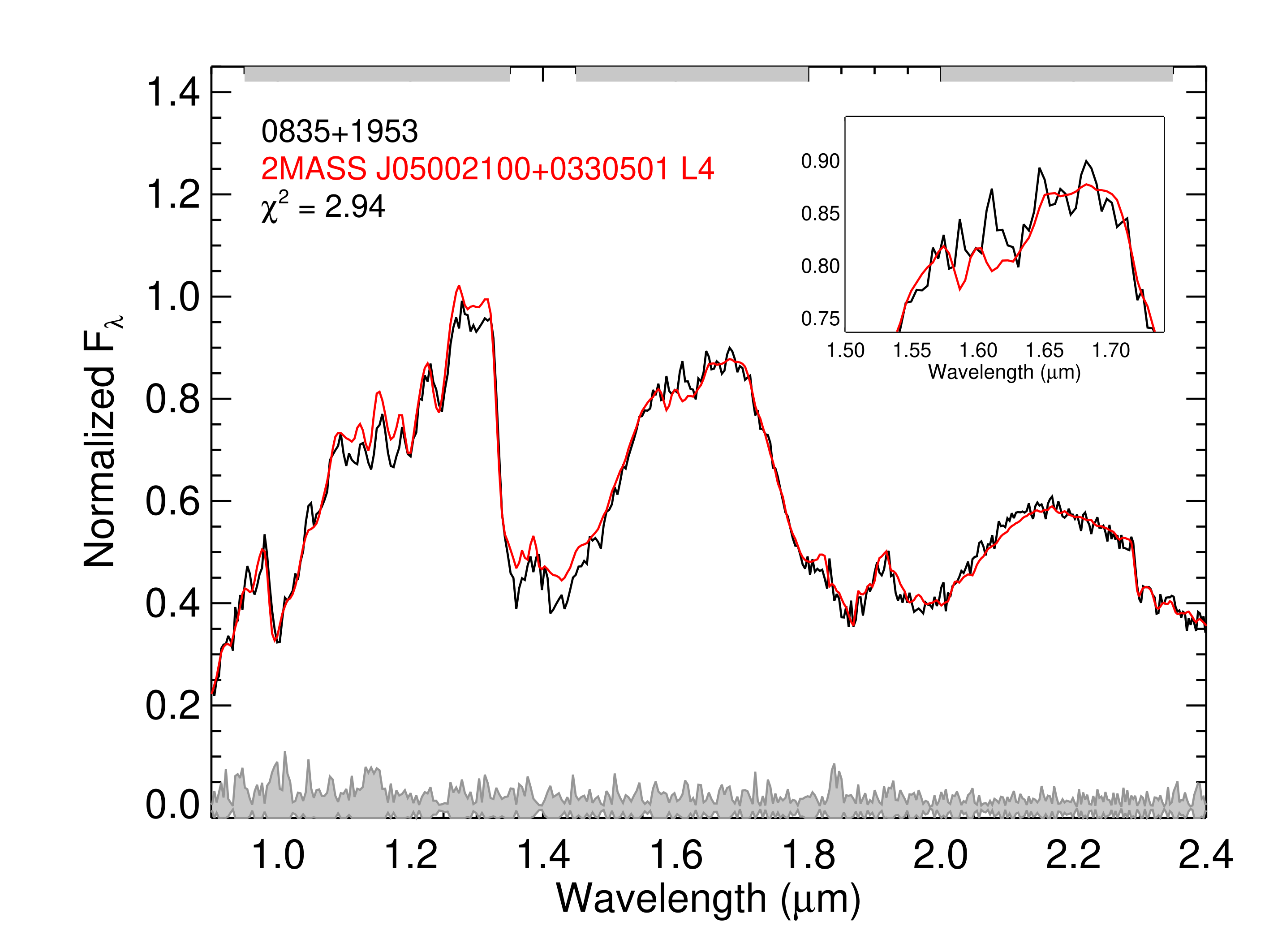}
  \end{minipage}
  \hfill
  \begin{minipage}[b]{0.49\textwidth}
    \includegraphics[width=\textwidth]{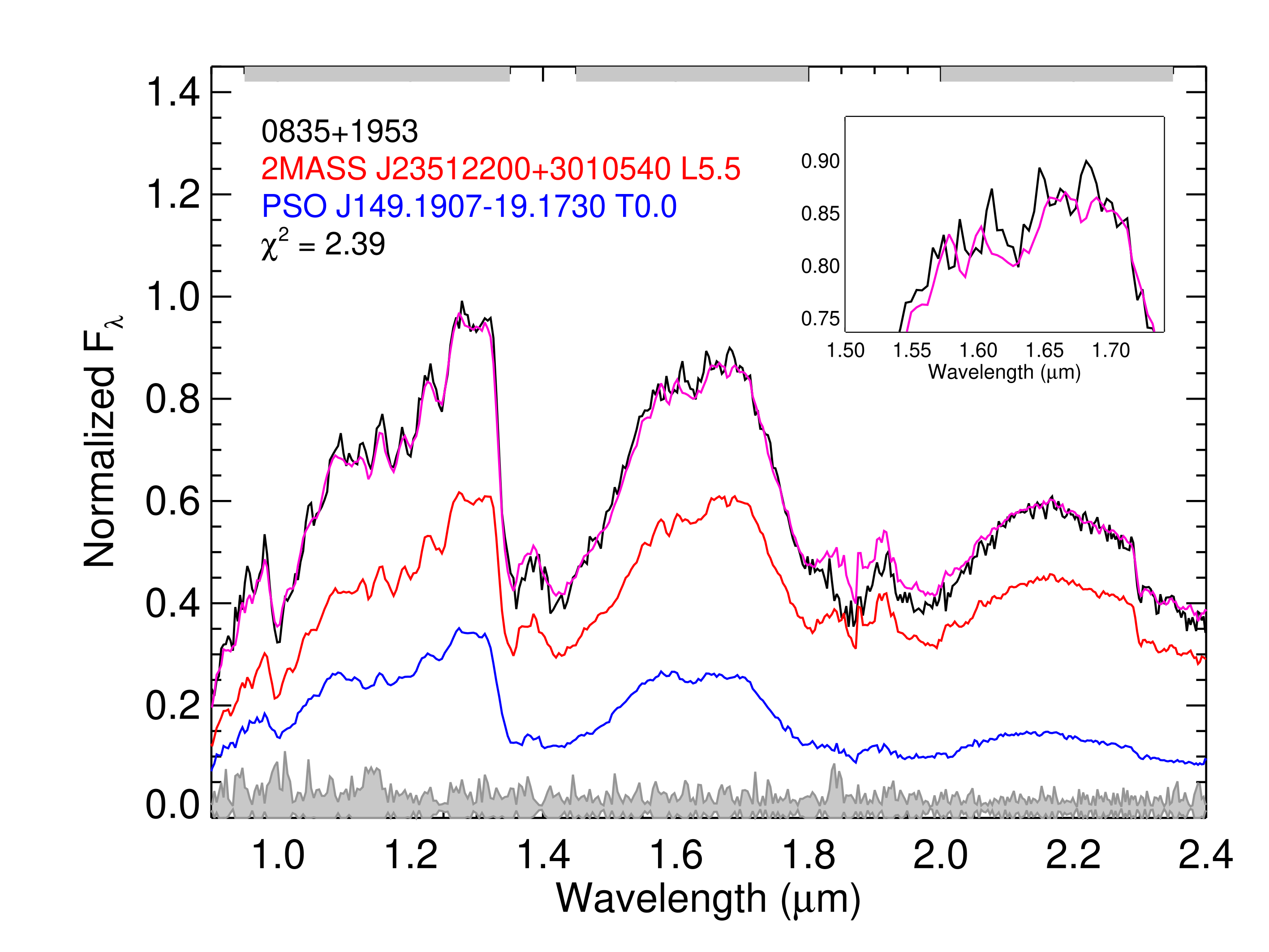}
  \end{minipage}
  \caption{Left: 2MASS J0835+1953's SpeX Prism spectrum (black) along with the best single spectral template fit (red). Grey boxes along the top axis denote the wavelength ranges of the $J$-, $H$-, and $K_S$-bands. The shaded region along the lower axis shows the uncertainty on 2MASS J0835+1953's spectrum as a function of wavelength. The inset shows a zoom-in on a region surrounding a potential absorption feature from CH$_4$ at $1.63$ $\mu$m, which may indicate the presence of a T dwarf companion in the spectrum. Right: the best fitting binary template (magenta), which consists of a L5.5 primary (red) and a T0.0 secondary (blue).}
  \label{fig:binary_fits}
\end{figure*}

We show the best-fitting single and binary templates in Figure \ref{fig:binary_fits}. Of particular note is an absorption feature near $1.63$ $\mu$m (shown in detail in the inset of each panel), which is indicative of methane absorption \citep[e.g.,][]{Burgasser2007} and hints at the presence of a T dwarf companion hidden in the spectrum. However, the low SNR of the spectrum prevents us from determining at present whether this feature is truly due to $CH_4$ absorption or due to noise. In summary, while our results suggest that 2MASS J0835+1953's SpeX Prism spectrum is most likely attributable to a single object, we cannot definitively rule out the possibility that it is a binary system, and to do so will require a higher SNR spectrum.

\subsubsection{Follow-up Observations}

\begin{deluxetable*}{lllrrrrl}
        \tablecaption{Follow-up observations of 2MASS J0835+1953. As in Table \ref{tab:observations}, a standard deviation is provided for data taken with each unique exposure time.}
        \tablenum{3}
        \label{tab:followup}
        \tablehead{
            \nocolhead{}      & \colhead{Facility/}  & \nocolhead{}     & \colhead{Duration}          & \colhead{Exp.} & \nocolhead{}       & \colhead{Seeing} & \nocolhead{}\\ 
            \colhead{UT Date} & \colhead{Instrument} & \colhead{Filter} & \colhead{(hr)} & \colhead{Time (s)} & \colhead{$\sigma$} & \colhead{($^{\prime\prime}$)} & \colhead{Notes}
            }
            
        \startdata
            2021 Dec. 12 & PTO/Mimir & MKO $J$ & 7.4 & 60 & 0.052 & 2.4 &\\
            2021 Dec. 13 & PTO/Mimir & MKO $J$ & 7.6 & 60 & 0.052 & 2.4 &\\
            2021 Jan. 13 & PTO/Mimir & MKO $J$ & 8.7 & 60 & 0.054 & 2.3 &\\
            2022 Jan. 15 & PTO/Mimir & MKO $J$ & 8.8 & 60 & 0.066 & 3.0 &\\
            2022 Jan. 20 & PTO/Mimir & MKO $J$ & 8.9 & 30, 60 & 0.075, 0.061 & 2.4 &\\
            2022 Jan. 21 & PTO/Mimir & MKO $J$ & 8.9 & 60 & 0.042 & 2.3 &\\
            2022 Feb. 10 & LDT/LMI   & SDSS $z$ & 4.1 & 60 & 0.016 & 2.4 & Second half lost to wind\\
            2022 Feb. 11 & PTO/Mimir & MKO $J$ & 8.9 & 60 & 0.062 & 2.5 & Clouds first two hours\\
            2022 Feb. 12 & PTO/Mimir & MKO $J$ & 8.9 & 60 & 0.050 & 2.5 &\\
            2022 Feb. 13 & PTO/Mimir & MKO $J$ & 8.9 & 60 & 0.052 & 2.5 &\\
            2022 Feb. 14 & PTO/Mimir & MKO $J$ & 8.9 & 60 & 0.051 & 2.4 &\\
            2022 Feb. 18 & PTO/Mimir & MKO $J$ & 8.7 & 30, 60 & 0.130, 0.060 & 2.6 & High background first hour\\
            2022 Feb. 19 & PTO/Mimir & MKO $J$ & 8.8 & 30, 60 & 0.123, 0.052 & 2.5 & High background first hour \\
            2022 Feb. 20 & PTO/Mimir & MKO $J$ & 8.5 & 30, 60 & 0.067, 0.041 & 2.3 & High background first hour\\
        \enddata
        
    \tablecomments{The seeing column gives the average seeing FWHM measured over the duration of the observations.}
\end{deluxetable*}

Unable to explain 2MASS J0835+1953's candidate transit event on the basis of source variability, a variety of diagnostic checks, or with a background eclipsing binary scenario, we began a dedicated follow-up campaign to search for a repeat transit. The observations are summarized in Table \ref{tab:followup}. We obtained 14 staring light curves, 13 with the PTO/Mimir in MKO $J$-band and 1 with the Large Monolithic Imager (LMI) on the 4.3-m Lowell Discovery Telescope (LDT) in SDSS $z$-band. In total, we have dedicated 116 hours to follow-up observations of 2MASS J0835+1953. Photometry was performed using time-variable apertures with radii set to $1\times$ a smoothed trend of the measured seeing FWHM values in each image.  

\begin{figure*}
    \centering
    \includegraphics[width=\textwidth]{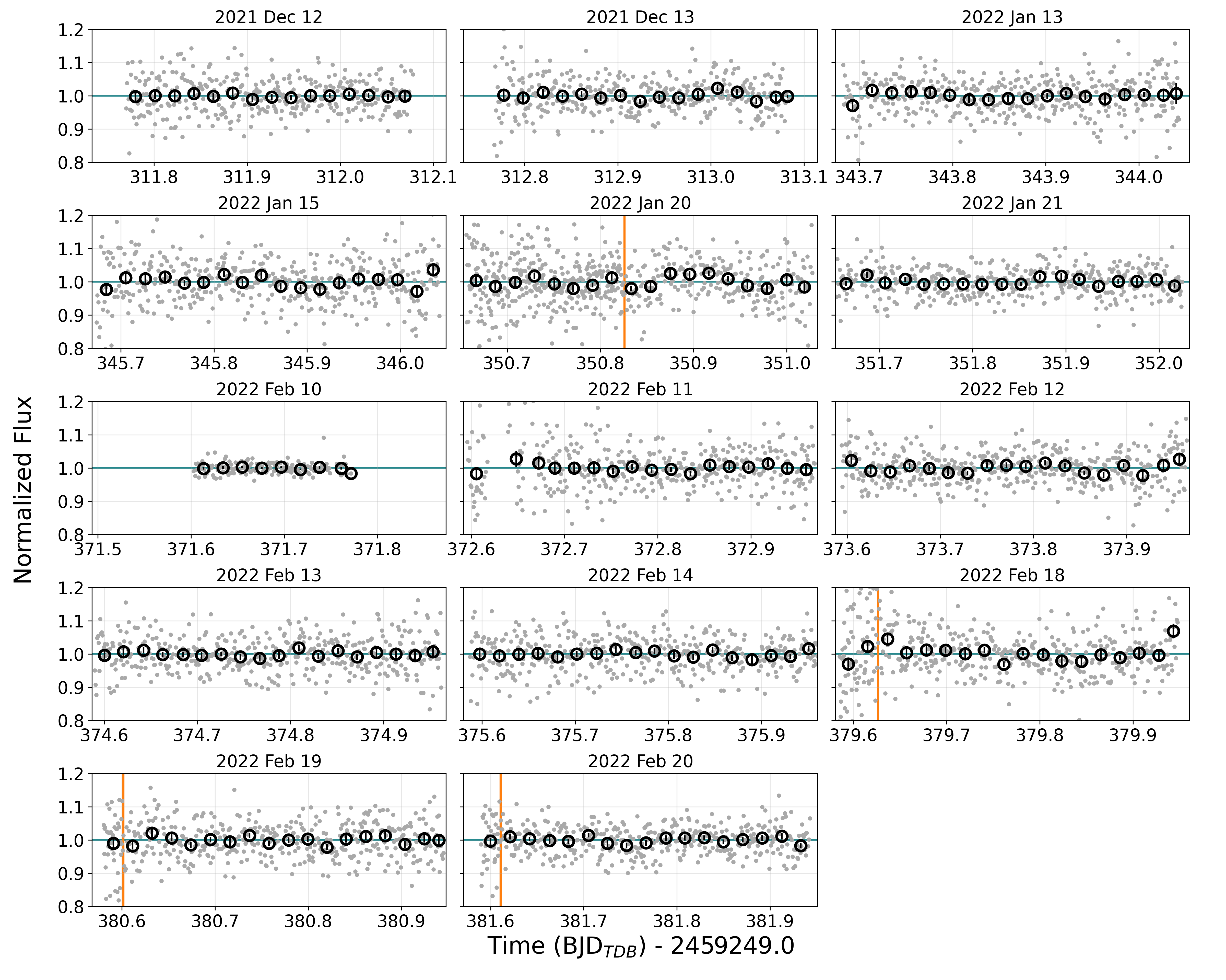}
    \caption{Follow-up light curves of 2MASS J0835+1953. The black points with error bars show the flux binned over 30 minute intervals. Each panel has the same time extent. An orange line marks the transition from 30 s to 60 s exposures in the appropriate light curves. A repeat transit event has not been observed, but these observations rule out source variability as a likely cause as the candidate event in Figure \ref{fig:2m0835}.}
    \label{fig:followup}
\end{figure*}

The follow-up light curves are shown in Figure \ref{fig:followup}. The black points with error bars show the flux measurements binned over 30 minute intervals, which is about half the expected duration of a transit around a typical L or T dwarf. These data have the sensitivity to detect a repeat $8.9\%$ transit event, with an average 30-minute $\sigma$ of 1.1\%, but a repeat transit was not observed.

\begin{figure}
    \centering
    \includegraphics[width=\columnwidth]{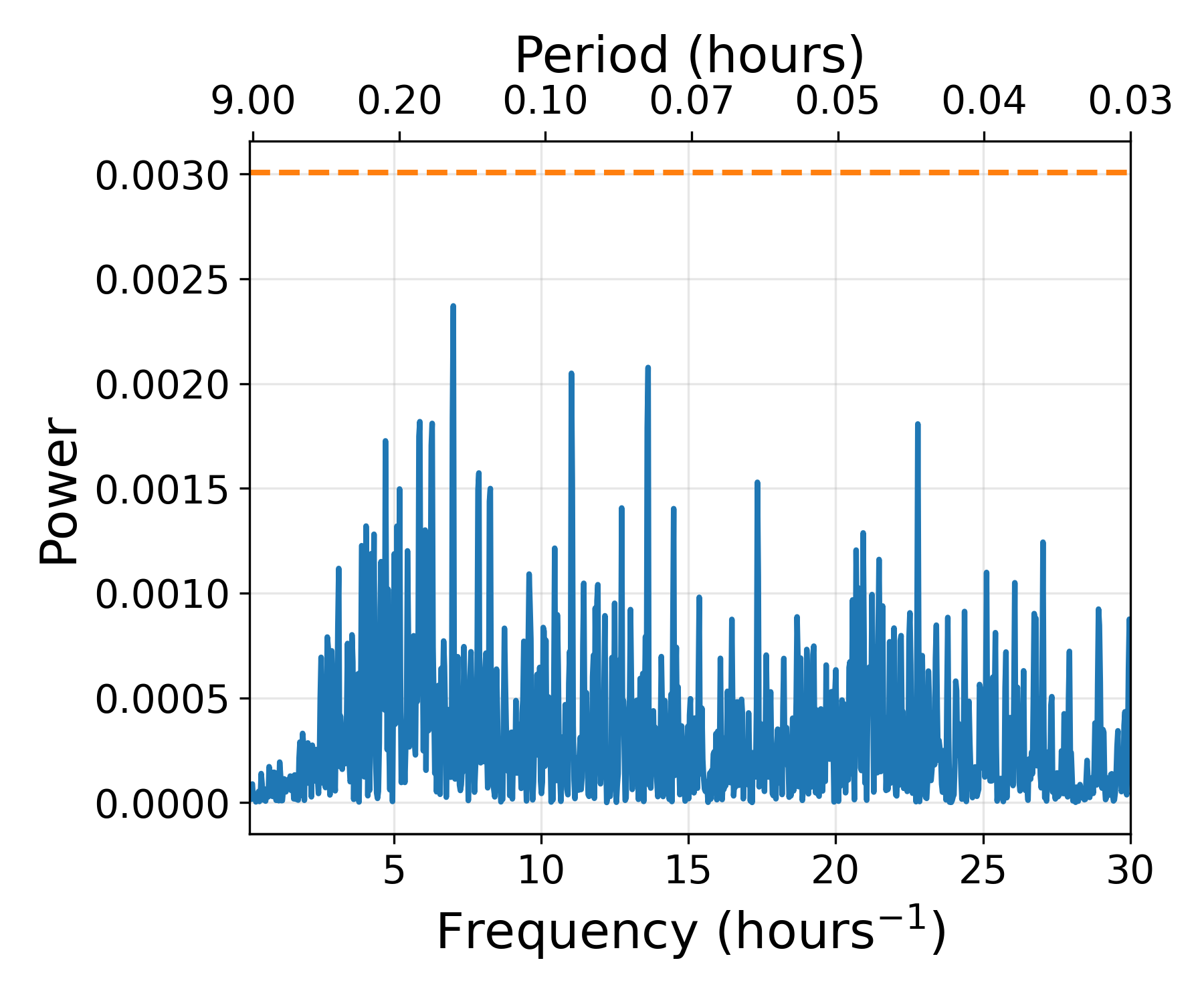}
    \caption{A Lomb-Scargle periodogram of all the unbinned data collected for 2MASS J0835+1953. A dashed orange line shows the 95\% false alarm probability threshold. No significant peaks were detected in the periodogram, or in periodograms constructed using individual nights of data.}
    \label{fig:periodogram}
\end{figure}

However, these observations strongly diminish the possibility that the candidate event in Figure \ref{fig:2m0835} is attributable to source variability. A Lomb-Scargle periodogram of the data, shown in Figure \ref{fig:periodogram}, revealed no significant periodicity over a frequency range of $0.111-30$ hr$^{-1}$ (corresponding to periods between two minutes and nine hours). This range was chosen to test for variability on timescales from twice the median exposure time up to the duration of the longest light curves. Since the light curves are normalized on a nightly basis, we are unable to test for variability on longer timescales; however, long-timescale variability is irrelevant for testing for periodicity that could explain a $\sim1$ hr transit event. The highest peak in the periodogram has a false alarm probability of $1.0$, indicating that no significant variability was detected. We also tested for variability on a night-by-night basis, creating a Lomb-Scargle periodogram for each night of data with period coverage ranging from two minutes up to the duration of each night's light curve. No significant peaks were found in any of the single-night periodograms.

\subsubsection{MCMC Modeling}
\label{sec:mcmc}
With source variability, known systematics, and a background EB scenario ruled out, a transiting companion remains as a viable explanation for the flux dip in 2MASS J0835+1953's light curve. We therefore performed a Markov chain Monte Carlo (MCMC) simulation to estimate the values and uncertainties of the transit parameters, using \texttt{emcee} \citep{emcee2013} to sample the posteriors of \texttt{BATMAN} \citep{Kreidberg2015} transit light curve models. We performed this sampling using all of the unbinned data obtained for 2MASS J0835+1953 (i.e., all of the grey points in Figures \ref{fig:2m0835} and \ref{fig:followup}). We placed a Gaussian prior on the mid-transit epoch $T_0$ with a mean equal to $2459249.9093$ (the mid-point of the alleged in-transit block) and a standard deviation of $0.015$ days. This prior allows for models with $T_0$'s between the two surrounding blocks of data, as we cannot be sure that our light curve sampled the true mid-transit time. 

We sampled values for the orbital period $P$ in days in $\log_{10}$ space, with a uniform prior between $0$ and $2$. Log sampling was used to discourage solutions with planets on long orbital periods, which have a lower geometric transit probability ($\propto P^{-2/3}$) and a lower probability of transiting during our observations ($\propto P^{-1}$). Several studies of single transit light curves have applied informative priors on $P$ that can be used to place stronger constraints on the orbital period \citep[e.g.,][]{ForemanMackey2016,Osborn2016,Uehara2016,Kipping2018}; however, these priors were developed using continuous data sets from Kepler and K2, whereas PINES data are not continuous. 


We used another Gaussian prior on the orbital inclination $i$, with a mean of $90^\circ$ and a standard deviation of $5^\circ$ to allow for grazing transits. Eccentricity $e$ was locked to zero and a quadratic limb darkening model was assumed, with coefficients $u_1$ and $u_2$ locked to their expected values for a 2000 K $\log{g}=5.0$ host in $J$-band \citep{Claret2011}. We applied a uniform prior on the density of the host, $\rho_h$, between 40 g cm$^{-3}$ and 60 g cm$^{-3}$. These bounds were calculated using radii and masses from the evolutionary models of \citet{Baraffe2015} for ages $> 0.1$ Gyr. Finally, we placed a uniform prior on the planet-to-host radius ratio, $R_p/R_h$, which was allowed to range from $0-1$. With an incomplete transit event, we cannot assume that our light curve samples the deepest in-transit point. This prior allows for an eclipsing brown dwarf scenario by permitting companion radii up to the size of the host.

\begin{figure}
    \centering
    \includegraphics[width=\columnwidth]{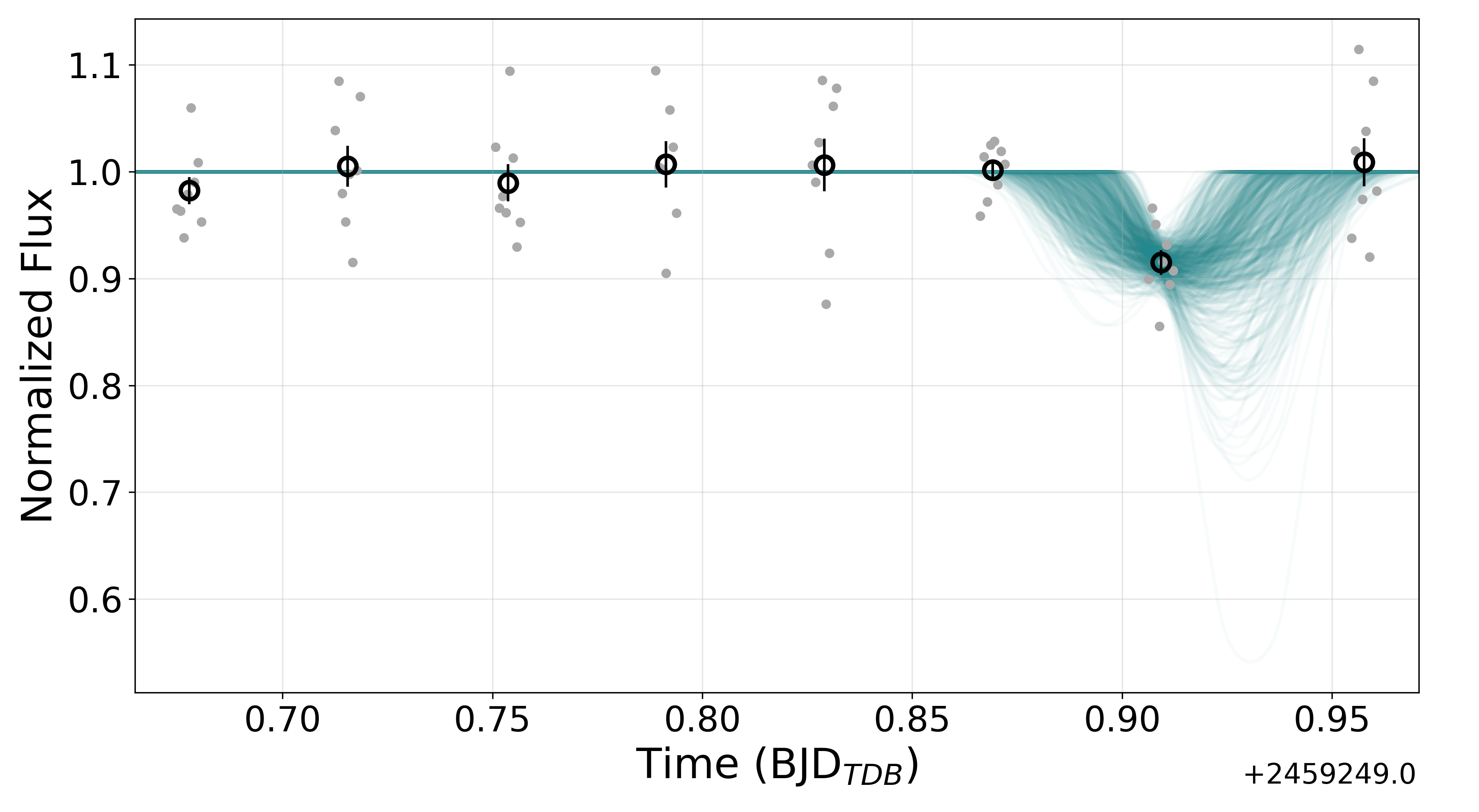}
    \caption{Random samples from the MCMC on the night of the candidate transit. Note that the sampling was performed using all of 2MASS J0835+1953's unbinned data, and that we are only showing the night of the candidate transit for clarity. Our data is consistent with two general classes of solutions: one in which our observations are the deepest part of the transit event, and one in which we sampled the ingress of a deeper event.}
    \label{fig:mcmc}
\end{figure}

\begin{figure*}
    \centering
    \includegraphics[width=\textwidth]{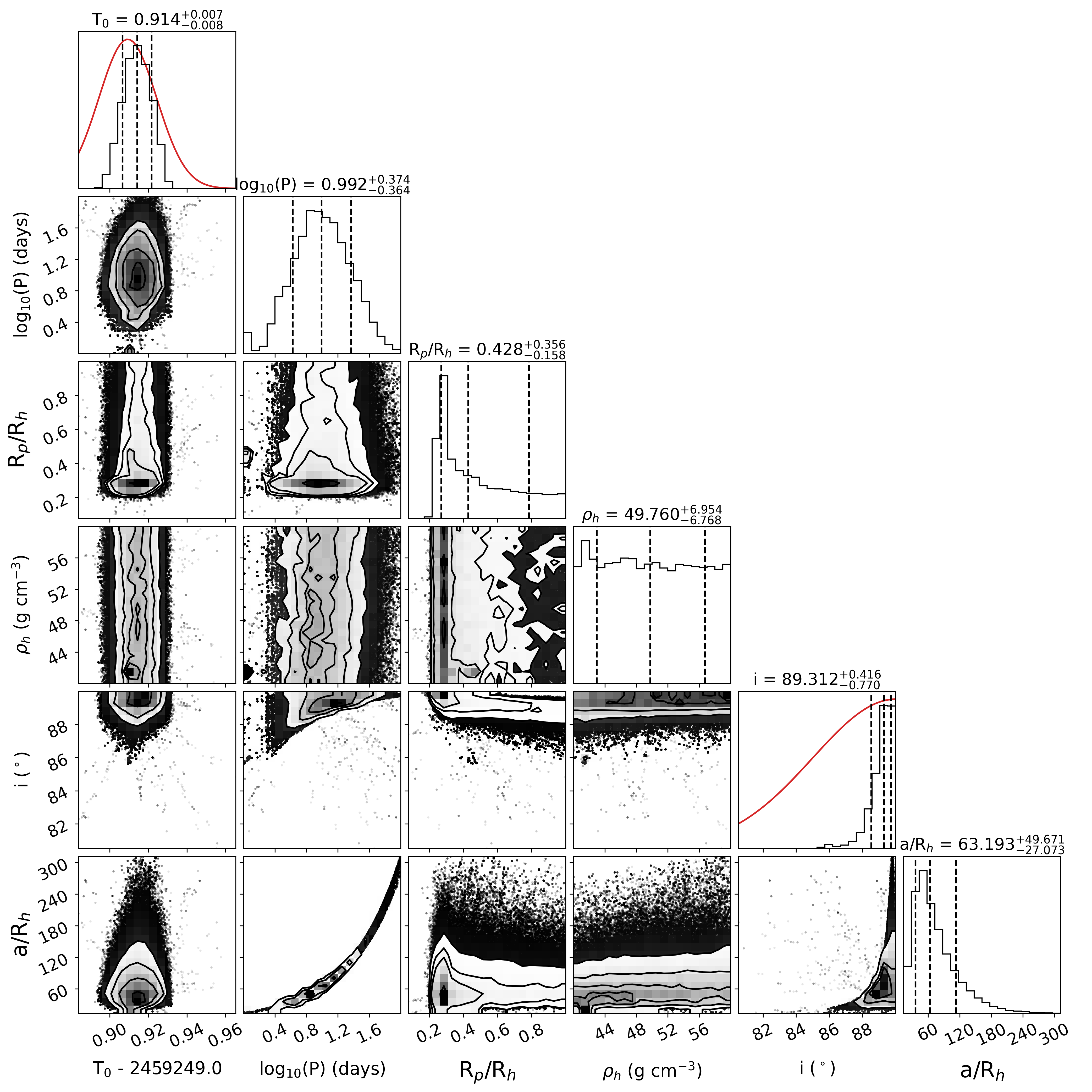}
    \caption{A corner plot of the parameters in the MCMC simulation, with the 1$\sigma$ range of each parameter indicated with dashed lines. Red curves show the Gaussian priors on $T_0$ and $i$. The $a/R_h$ posterior was calculated using the $P$ and $\rho_h$ posteriors with Kepler's third law. This plot was made with \texttt{corner} \citep{corner2016}.}
    \label{fig:corner}
\end{figure*}

We ran 64 chains for $20,000$ steps, with the first 2,000 steps discarded as burn-in. Figure \ref{fig:mcmc} shows the night of the candidate event with random samples from the MCMC plotted as faint lines. This plot shows that our data is generally consistent with two clusters of solutions: one where the alleged in-transit block is the deepest part of the transit, and one where our observations sampled the ingress of a deeper event. The posterior samples tend to disfavor an egress scenario because there is a slight downward slope to the points during the suspected in-transit block. Figure \ref{fig:corner} shows a corner plot of the parameters that were allowed to vary in the MCMC. The 15.9\%, 50\%, and 84.1\% quantiles of each parameter are shown as dashed lines. We find that $T_0=2459423.914^{+0.007}_{-0.008}$ BJD$_{TDB}$, $R_p/R_h=0.428^{+0.356}_{-0.158}$, and $i=89.31^{+0.42}_{-0.77}$ degrees. The orbital period is not well constrained by our observations, but the solutions do disfavor periods in the range of $\sim1.5\textendash2$ days. Figure \ref{fig:corner} also includes the posterior distribution of $a/R_h$, which was calculated using the $P$ and $\rho_h$ posteriors along with Kepler's third law. We find a median $a/R_h$ of $63.2^{+49.7}_{-27.1}$.

\begin{figure}
    \centering
    \includegraphics[width=\columnwidth]{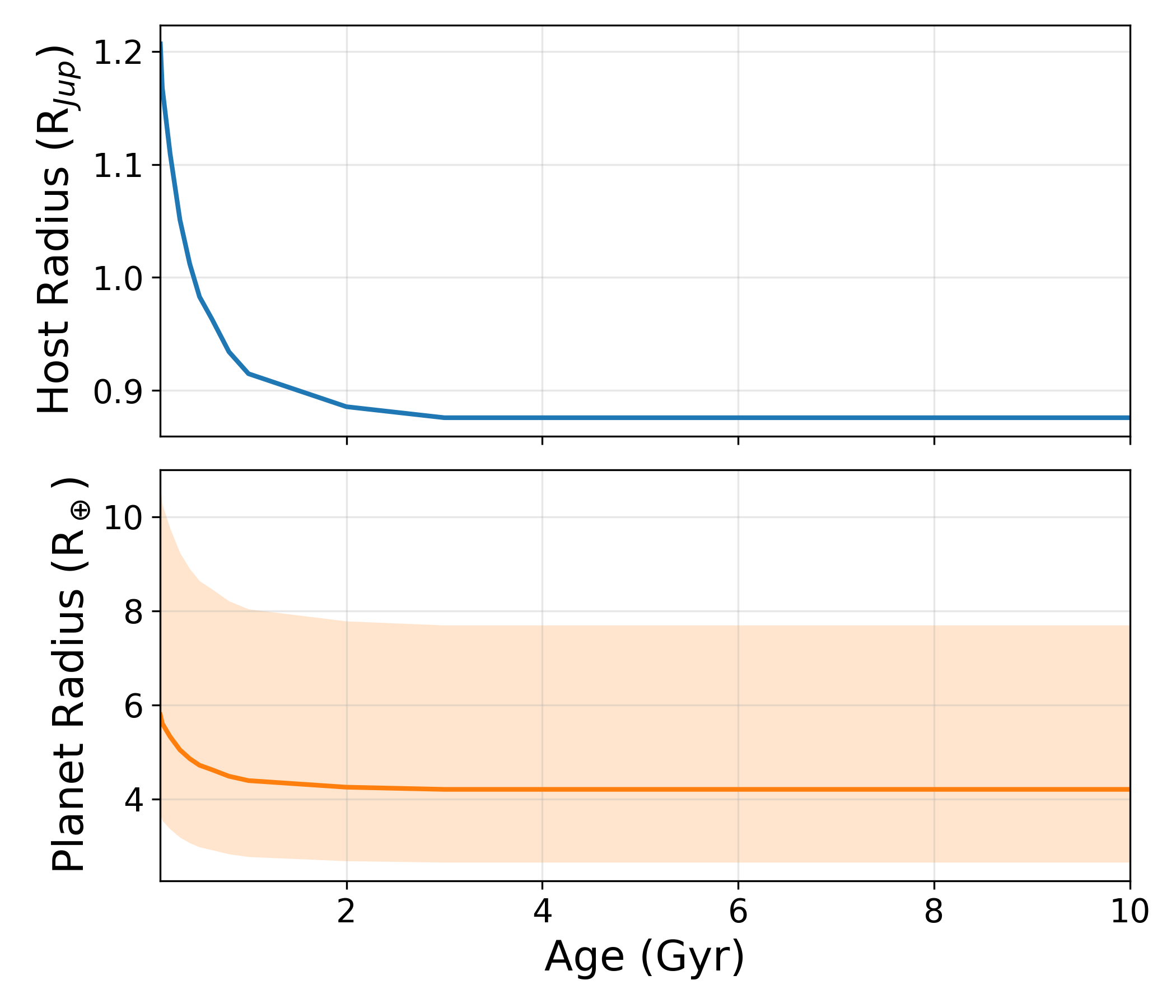}
    \caption{Top: the radius that an L5 ($T_{eff}\approx1600$ K) brown dwarf would have depending on its age over the range $0.1-10$ Gyr \citep{Baraffe2015}. Bottom: the corresponding radius of the planet candidate based on the $R_p/R_h$ value from our MCMC modeling. The shaded region indicates the $1\sigma$ uncertainty on the planet radius.}
    \label{fig:radius_age}
\end{figure}

Brown dwarfs contract considerably as they age and cool. As a consequence, our estimate of the physical radius of the planet candidate is age dependent. The field surface gravity classification of 2MASS J0835+1953 suggests an age $\gtrsim0.1$ Gyr \citep{Martin2017}, but this does not strongly constrain its radius, a fact that we visualize in the top panel of Figure \ref{fig:radius_age}. This panel shows the radius that an L5 ($T_{eff}\approx1600$ K) brown dwarf would have over a range of $0.1-10$ Gyr in age using the evolutionary models of \citet{Baraffe2015}. Note that this does not depict the evolutionary path of 2MASS J0835+1953, because brown dwarfs assume progressively later spectral types as they age and cool. Rather, it shows the radius that a 1600 K object would have depending on its actual age. If 2MASS J0835+1953 is very young, it could have a radius as large as $1.21$ $R_{Jup}$; if it is older and has fully contracted, its radius is as small as $0.88$ $R_{Jup}$. From a frequentist perspective, a smaller radius is more likely: if one assumes random ages drawn from a uniform distribution from $0.1-10$ Gyr, they would identify the fully contracted radius in 76\% of cases. However, we emphasize that we cannot rule out larger radii without placing firmer constraints on the target's age. 

The bottom panel of Figure \ref{fig:radius_age} shows the effect of contraction translated into the radius estimate of the candidate planet using the $R_p/R_h$ value from our MCMC simulation. If the host is very young, our simulation suggests a planet radius as large as \upperrad; if the host has fully contracted, we find a planet radius of \lowerrad.

\section{Injection and Recovery Simulation} \label{sec:injection_recovery}
Kepler discoveries have established a significant increase in the occurrence rate of short-period Earth- to Neptune-sized exoplanets from F- to M-type stars \citep{Howard2012, Dressing2015, Mulders2015a,Mulders2015b,HardegreeUllman2019}. All things being equal, and assuming no significant differences between the formation mechanism of the lowest-mass stars (e.g., TRAPPIST-1) and brown dwarfs, it is not unreasonable to expect that this trend could continue into the L and T spectral types. As a simple test of this hypothesis, we performed an injection and recovery simulation.

We used the 131 PINES light curves summarized in Table \ref{tab:observations} in the simulation, masking out the candidate transit events in the light curves of 2MASS J1821+1414 and 2MASS J0835+1953. Planets were injected using measured M dwarf short-period planet occurrence rates from \citet{Dressing2015}, which were calculated over a grid in planet radius versus orbital period space using Kepler discoveries. 

For every light curve in our sample, we first assigned the target a random on-sky inclination $i$ drawn from a $\sin{i}$ distribution. We used each target's spectral type to estimate an effective temperature using the the $T_{eff}$ -spectral-type relation for field M6-T9 dwarfs from \citet{Faherty2016}, and assigned each a random age from $0.005-10$ Gyr. We used the age and $T_{eff}$ values to assign each target mass ($M_h$) and radius ($R_h$) values from the nearest evolutionary model point from \citet{Baraffe2015}. 

Then, for each target, we looped over all cells in the occurrence rate grid and drew a random number from a uniform distribution between zero and one. If this number was less than the occurrence rate in the grid cell, we proceeded with the planet injection, drawing a $P$ and $R_p$ within the bounds of the given cell. We assigned a random $T_0$ by drawing a random number in the range $0-P$ days, and calculated the orbital semi-major axis $a$ assuming circular orbits (i.e., $e = 0$). We generated \texttt{BATMAN} transit models \citep{Kreidberg2015} using the random planet parameters and assuming quadratic-limb darkening coefficients for a 2000 K target with $\log{g}=5.0$ in \textit{J}-band \citep{Claret2011}. We then injected the planet by multiplying the light curve in question by the transit model. 

Next, we applied our detection algorithm (Section \ref{sec:transit_search_and_recovery_algorithm}) to the planet-injected light curve. We considered a planet to be detected if our algorithm correctly flagged the appropriate block of data as containing a transit candidate with an SNR$'$ greater than six. We repeated this process for all of the occurrence rate grid cells and for all of the light curves in the sample. 

\begin{figure*}
    \centering
    \includegraphics[width=\textwidth]{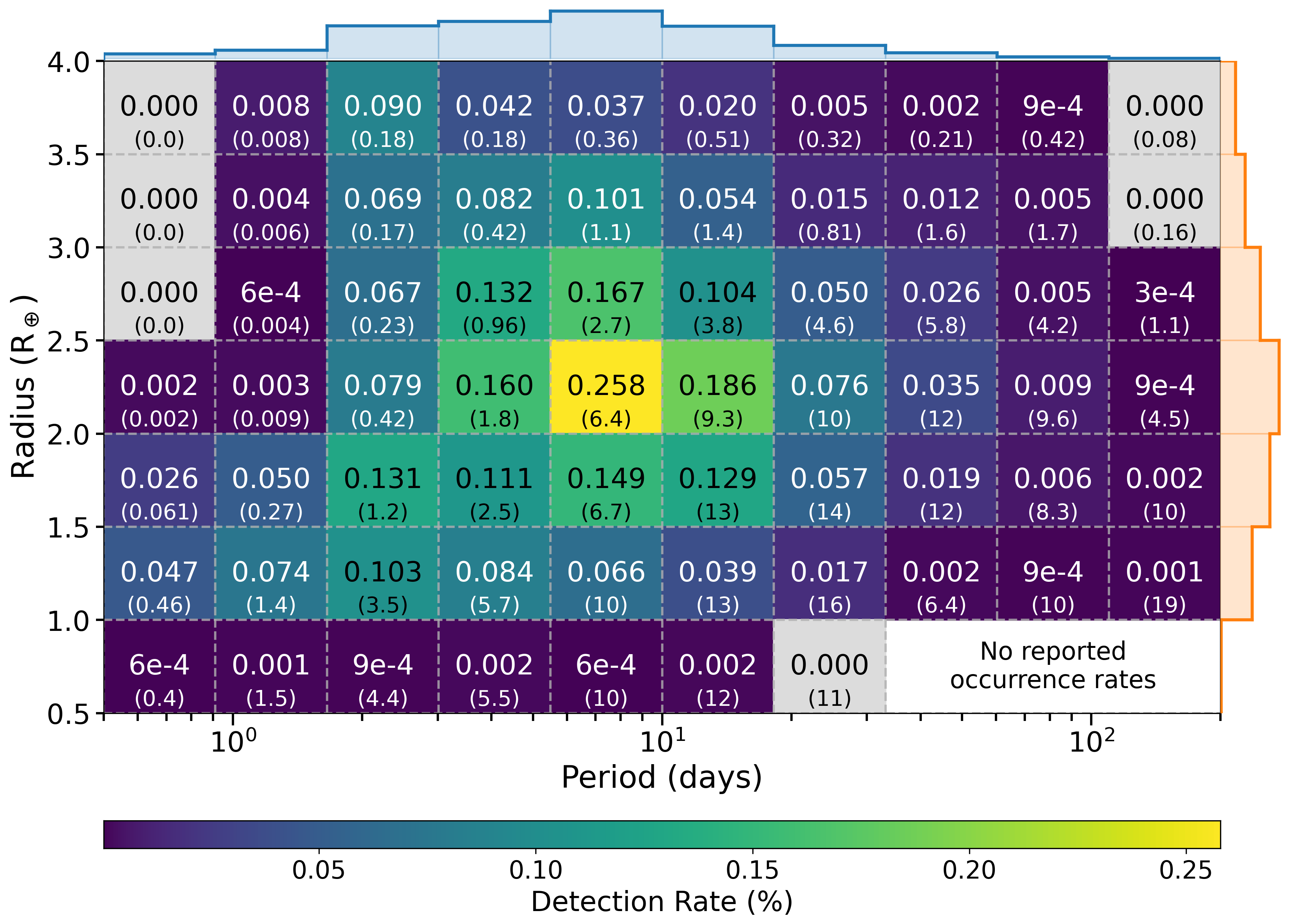}
    \caption{The results of our injection and recovery simulation. Grid cells are colored by their detection rate in percent, which is also listed in the center of each cell. The occurrence rate of planets in each grid cell, as calculated in \citet{Dressing2015}, is given in parentheses at the bottom of each cell (also in percent). Cells in which no planets were recovered are colored in grey. Histograms along the top and right axes show the summed detection rates along the period and radius dimensions, respectively.}
    \label{fig:injection}
\end{figure*}

We performed this simulation repeatedly, with each iteration representing the acquisition of 131 PINES light curves injected with a random population of M-dwarf-style short-period planets. We ran the simulation until $10,000$ total planets were detected, which required $329,870$ iterations. By dividing the total number of detections in each grid cell by the number of iterations, we calculated a map of expected detection rates, which we show in Figure \ref{fig:injection}. Grid cells are colored by their detection rate (in percent), which is also reported in the center of each cell. The corresponding occurrence rate from \citet{Dressing2015} is listed at the bottom of each cell in parentheses. Cells in which no detections were made are colored in grey. 

This figure reveals a complex interplay between the assumed planet occurrence rates, transit probability, transit observability, and detection completeness. The simulation suggests that PINES observations are most sensitive to detecting planets with periods from $\sim1.7-18$ days and radii from $\sim1.5-3.0$ $R_\oplus$. While super-Earth planets with periods greater than $18$ days occur frequently around M dwarfs, they should rarely be detected in PINES data. This is because such planets have lower transit probabilities, and, if they do transit, are likely to not be observed, with $T_0$ values that lie outside of the limited time coverage of PINES light curves. While large planets with short orbital periods are easy to detect, they are not frequently recovered in our simulation, owing to their intrinsic rareness around M dwarfs. 

Summing the detection rates of individual grid cells in Figure \ref{fig:injection} gives the total detection rate of planets of all radii and periods. We calculate this number to be $3.03$\%, indicating that under the assumption of M dwarf short-period planet occurrence rates, we should have a $3.03$\% total chance of detecting a transit signature in our 131 light curves using our detection algorithm. 

We can also use this map to estimate the likelihood of detecting a planet with a radius similar to the one estimated for 2MASS J0835+1953 b. As discussed in Section \ref{sec:mcmc}, the candidate's physical radius is dependent on the age of the host, and may be as large as \upperrad if the host is young, or as small as \lowerrad if the host has fully contracted. We therefore performed two calculations corresponding to the lower $1\sigma$ bound on both of these limits. If the host has fully contracted, the lower $1\sigma$ bound on 2MASS J0835+1953 b is $2.6 R_\oplus$; we calculate a $1.00\%$ total chance of detecting a planet with this radius or larger in our injection and recovery simulation. If the host is young and the lower $1\sigma$ limit on the planet radius is instead $3.7 R_\oplus$, this number changes to $0.13\%$. 

Regardless of the true age of 2MASS J0835+1953, these results suggest that we are unlikely to detect a planet with the necessary radius in our collection of 131 light curves unless the occurrence rates of short-period planets around L and T dwarfs are higher than those around M dwarfs. Thus, if the planet candidate were confirmed, it would be suggestive of an enhancement in the occurrence rate of such planets around L and T dwarfs compared to M dwarfs. Entertaining this possibility for the sake of argument, an enhanced occurrence rate of short-period super-Earths around L and T dwarfs would challenge current planet formation models, which struggle with the formation of super-Earths around these spectral types \citep{Payne2007, Miguel2020, Mulders2021}. However, even if the planet candidate is confirmed, more detections would be required to determine whether its discovery was a function of luck or of truly enhanced occurrence rates. 

\section{Conclusions} \label{sec:conclusions}
We described a new transit detection algorithm that is designed to detect transit events in PINES data. Our observing strategy biases us heavily toward the detection of single, incomplete transits, and the L and T spectral types which compose our sample frequently exhibit significant, short-timescale photometric variability due to rotation coupled with heterogeneous surface features. Our detection algorithm uses a QP-GP model to search for significantly decremented blocks of data in potentially variable light curves. 

We used this algorithm to search for transits in a collection of 131 PINES light curves. We identified two candidate events, one in the light curve of 2MASS J1821+1414 and the other in the light curve of 2MASS J0835+1953. We disfavor a transiting planet scenario for explaining 2MASS J1821+1414's candidate event because the object is a known irregular variable with observed percent-level variability in $J$-band. We could not rule out a transiting planet in the case of 2MASS J0835+1953 with known source variability properties or with a variety of diagnostic checks. 

We performed a dedicated follow-up campaign of 2MASS J0835+1953 to try and recover a second transit. We obtained 116 hours of data on the target and have yet to observe a repeat flux dimming event. However, these observations also revealed that the source does not exhibit significant variability over two minute to nine hour timescales, which strongly disfavors another potential explanation of its candidate transit event. We performed an MCMC sampling of 2MASS J0835+1953's light curve, and estimated a planet radius between \lowerrad and \upperrad, depending on the age of the host.

Finally, we performed an injection and recovery simulation using our light curve sample and transit detection algorithm. Planets were injected into the data using measured M dwarf short-period planet occurrence rates. The results of this simulation suggested a $\sim1\%$ chance of detecting a planet candidate like 2MASS J0835+1953 b in our data set \textit{if} M dwarf short-period planet occurrence rates hold for L and T dwarfs. If 2MASS J0835+1953 b is confirmed, it would therefore be suggestive an enhancement in the occurrence of such planets around L and T dwarfs, and would challenge findings from models of planet formation around sub-stellar objects. However, additional planet detections around these spectral types would have to be made to confirm a true enhancement in the occurrence rates. 

2MASS J0835+1953 b joins 2MASS J1119–1137 AB b \citep{Limbach2021} as the only reported planet candidates around brown dwarfs, to date. We intend to continue our follow-up campaign of 2MASS J0835+1953, but the effectiveness of this effort will be limited by the typical constraints of ground-based observing, namely weather losses and gaps during the day. However, with the loss of Spitzer, there remain limited suitable options for continuous space-based follow-up. 

For example, while both 2MASS J1821+1414 and 2MASS J0835+1953 have been observed with the Transiting Exoplanet Survey Satellite \citep[TESS,][]{Ricker2015}, the bandpass  ($\sim600-1100$ nm) and small aperture ($10.5$ cm) of this facility renders it unsuitable for confirming the planetary nature their light curves. We scaled BT-Settl model spectra \citep{Allard2012} for each object using their measured 2MASS $J$-band magnitudes, and calculated TESS magnitudes of $17.0$ and $19.7$, respectively. In both cases, this equates to a one-hour standard deviation greater than 5\% \citep[estimated with \texttt{ticgen},][]{ticgen2017}. The prospects for space-based follow-up are unlikely to improve in the near future. While photometry with the James Webb Space Telescope would undoubtedly have the sensitivity to confirm 2MASS J0835+1953 b or 2MASS J1119–1137 AB b, searching for planets with unconstrained orbital periods would be a misallocation of that facility's limited time budget. Similar arguments could be made against an extended follow-up campaign with the oversubscribed Hubble Space Telescope, which would additionally have to contend with data discontinuities due to its low Earth orbit. Other planet detection techniques (radial velocity, direct imaging, or astrometry) will not be practical for confirming 2MASS J0835+1953 b with current facilities.

Ground-based surveys like PINES and SPECULOOS are demonstrating the ability to detect super-Earth-sized transiting exoplanets around ultracool dwarfs. Regardless of whether or not 2MASS J0835+1953 b is confirmed, these projects have the sensitivity to extend our knowledge of short-period exoplanet occurrence rates into the substellar regime, and will test competing predictions from occurrence rate trends and planet formation simulations.

\begin{acknowledgments}
    The authors thank their anonymous referee for a thorough review that improved the quality of this work.

    This material is based upon work supported by the National Aeronautics and Space Administration under Grant No. 80NSSC20K0256 issued through the Science Mission Directorate.
    
    Additional support was provided by NASA through the NASA Hubble Fellowship grant HST-HF2-51447.001-A awarded by the Space Telescope Science Institute, which is operated by the Association of Universities for Research in Astronomy, Inc., for NASA, under contract NAS5-26555.
\end{acknowledgments}

%

\vspace{5mm}
\facilities{Boston University Perkins Telescope Observatory (Mimir),
            Lowell Discovery Telescope (Large Monolithic Imager)
}


\software{ \texttt{BATMAN} \citep{Kreidberg2015}, \texttt{celerite} \citep{ForemanMackey2017},  \texttt{corner} \citep{corner2016}, \texttt{emcee} \citep{corner2016}, \texttt{PINES Analysis Toolkit} \citep{PINES_analysis_toolkit}, \texttt{ticgen} \citep{ticgen2017}, \texttt{scipy} \citep{SciPy-NMeth2020}
          }





\bibliography{biblio}{}
\bibliographystyle{aasjournal}



\end{document}